\documentclass[fleqn,usenatbib]{mnras}

\usepackage{newtxtext,newtxmath}

\usepackage[T1]{fontenc}

\DeclareRobustCommand{\VAN}[3]{#2}
\let\VANthebibliography\thebibliography
\def\thebibliography{\DeclareRobustCommand{\VAN}[3]{##3}\VANthebibliography}

\usepackage{graphicx}	
\usepackage{amsmath}	
\usepackage{hyperref}
\usepackage{natbib}
\usepackage{caption}
\usepackage{multirow}
\usepackage{geometry}
\usepackage{booktabs}
\usepackage[table]{xcolor}
\usepackage{array}
\usepackage{mathtools}
\usepackage{xspace}

\usepackage{multirow}

\newcommand{\reff}{\ensuremath{r_{\mathrm{e}}}\xspace}

\newcommand{\vcirc}{\ensuremath{v_{\mathrm{circ}}}\xspace}

\newcommand{\Msun}{\ensuremath{M_{\odot}}\xspace}

\newcommand{\Mstar}{\ensuremath{M_{\star}}\xspace}

\newcommand{\Mdyn}{\ensuremath{M_{\mathrm{dyn}}}\xspace}

\newcommand{\logMstar}{\ensuremath{\log(\Mstar ~\rm [\Msun])}\xspace}

\newcommand{\Ha}{\ensuremath{\mathrm{H}\alpha}\xspace}

\newcommand{\geko}{\textsc{geko}}

\newcommand{\sersic}{S\'{e}rsic}

\newcommand{\rotsupp}{\text{v}/\sigma_0} 
\newcommand{\disp}{\sigma_0}

\newcommand{\OIII}{$[\text{OIII}]$}
\newcommand{\PAmorph}{$\text{PA}_{\text{morph}}$}
\newcommand{\PAkin}{$\text{PA}_{\text{kin}}$}
\newcommand{\vre}{v_{\text{re}}}

\usepackage{adjustbox}  
\newcommand\orcid[1]{\href{http://orcid.org/#1}{\adjustbox{trim={-.15\width} {0\height} {-.15\width} {0\height},clip}{\includegraphics[height=10pt]{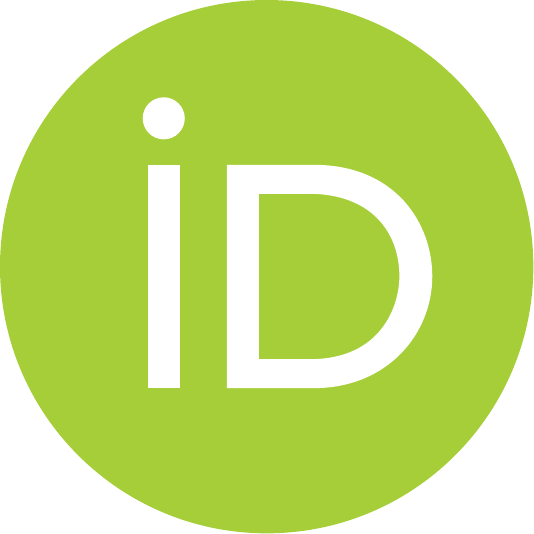}}}}

\title[\geko]{\geko: A tool for modelling galaxy kinematics and morphology in \textit{JWST}/NIRCam slitless spectroscopic observations}

\author[Danhaive \& Tacchella]{
A. Lola Danhaive\orcid{0000-0002-9708-9958}$^{1,2}$ \thanks{ald66@cam.ac.uk} and 
Sandro Tacchella\orcid{0000-0002-8224-4505}$^{1,2}$
\\
$^{1}$Kavli Institute for Cosmology, University of Cambridge, Madingley Road, Cambridge, CB3 0HA, UK \\
$^{2}$Cavendish Laboratory, University of Cambridge, 19 JJ Thomson Avenue, Cambridge, CB3 0HE, UK
}

\date{Accepted XXX. Received YYY; in original form ZZZ}

\pubyear{2025}

\begin{document}
\label{firstpage}
\pagerange{\pageref{firstpage}--\pageref{lastpage}}
\maketitle

\begin{abstract}
Wide-field slitless spectroscopy (WFSS) is a powerful tool for studying large samples of galaxies across cosmic times. With the arrival of \textit{JWST}, and its NIRCAM grism mode, slitless spectroscopy can reach a medium spectral resolution of $(R\sim 1600)$, allowing it to spatially resolve the ionised-gas kinematics out to $z\sim 9$. However, the kinematic information is convolved with morphology along the dispersion axis, a degeneracy that must be modelled to recover intrinsic properties. We present the Grism Emission-line Kinematics tOol (\textsc{geko}), a Python package that forward-models NIRCam grism observations and infers emission-line morphologies and kinematics within a Bayesian framework. \textsc{geko} combines \sersic\ surface-brightness models with arctangent rotation curves, includes full point-spread function (PSF) and line-spread function (LSF) convolution, and leverages gradient-based sampling via \textsc{jax}/\textsc{numpyro} for efficient inference. It recovers parameters such as effective radius, velocity dispersion, rotational velocity, rotational support, and dynamical mass, with typical run times of $\sim$20 minutes per galaxy on GPUs. We validate performance using extensive mock data spanning position angle, S/N, and morphology, quantifying where degeneracies limit recovery. Finally, we demonstrate applications to real FRESCO \Ha\ emitters at $z\approx4-6$, recovering both rotation- and dispersion-dominated systems. \textsc{geko} opens the way to statistical studies of galaxy dynamics in the early Universe and is publicly available at \url{https://github.com/angelicalola-danhaive/geko}.
\end{abstract}




\section{Introduction}
The dynamics of galaxies across cosmic time are a powerful probe of their evolutionary stages and mass assembly histories. Galaxies form through the accretion of gas into dark matter halos, where gas can cool to form stars.\citep{Fall:1980aa,Mo:1998aa}. During this process, the stars and the gas can be measured and carries crucial information on how the galaxies is building its stellar mass. This is particularly evident through disruptive events such as starburst driven outflows \citep{Oppenheimer:2008aa,Oppenheimer:2010aa,Brook:2012aa,Ubler:2014aa}, non-coplanar gas inflows, and mergers, which leave clear marks on the measured kinematics \citep{Sales:2012aa,Zolotov:2015aa,Tacchella:2019aa}.

Studies with integral field units (IFUs) from the local Universe out to Cosmic Noon ($z\approx 1-3$) have allowed for spatially resolved analyses of galaxy morphologies and kinematics. In particular, studies of ionised gas kinematics have uncovered the evolution of the star-forming galaxy population, at stellar masses $\log M_{\star}\rm ~[M_{\odot}]> 10-11$, from thick gas-rich disks at cosmic noon \citep[e.g.,][]{Wisnioski:2015vx, Price:2016uv, Tiley:2016aa, Simons:2017aa}, where the cosmic star formation rate (SFR) density peaks \citep{Madau:2014aa}, to thin cold disks in the local Universe \citep[e.g.][]{Epinat:2010aa}. This disk settling is often attributed to the decline of star formation and gas fractions with cosmic time. These IFU studies have also extended to stellar kinematics, using absorption lines, to characterize the kinematics of early-type galaxies \citep[see][ and references therein]{Cappellari:2016aa}.

The study of galaxy kinematics at high redshift was pioneered by the Atacama Large Millimeter/submillimeter Array (\textit{ALMA}), which probed cold gas (e.g. CO and [CII]) well into the Epoch of Reionization \citep[EoR; e.g.][]{Jones:2021aa, Bouwens:2022aa,Parlanti:2023ab,Zavala:2024aa,Scholtz:2025aa}. Although \textit{ALMA} enables novel and frontier science, it is restricted to targeted (and hence biased) observations and is not sensitive to low-mass galaxies \logMstar$\lesssim9$. With the arrival of \textit{JWST}, the warm ionized gas (e.g. \Ha\ and [OIII]) kinematics of galaxies have been explored beyond $z=4$ thanks to the NIRSpec/IFU \citep[e.g.,][]{Parlanti:2024aa, Jones:2024aa, Ubler:2024aa, Arribas:2024aa}. This instrument also allowed for the first constraints on resolved stellar kinematics beyond $z\approx 3$ \citep{DEugenio:2024ab,Belli:2024aa,Pascalau:2025aa}, crucial for studies of quiescent, or early-type, galaxies, which have no (or weak) emission lines. However, due to the IFU's limited field of view, this instrument typically focuses on single- or few-object studies. The NIRSpec micro-shutter assembly (MSA) is able to increase sample sizes with respect to the IFU, and has a point source resolution which can reach up to $R\approx 4000-5000$, subject to future calibration \citep{de-Graaff:2024ab}.

In contrast to these two instruments, NIRCam's grism mode offers the possibility of studying kinematics for large samples of galaxies,  This mode produces 2D spectra for all galaxies in the field-of-view (FoV), and is available in all of the NIRCam long wavelength (LW) filters, meaning it probes wavelengths $\lambda_{\rm obs} = 2.4-5.1 \thinspace \rm \mu m$. Thanks to its slitless nature, it does not suffer from the MSA's selection bias or partial slit covering, which is hard to quantify. The NIRCam grisms follow from their predecessors on the Hubble Space Telescope (HST)'s Wide Field Camera 3 (WFC3), which did not have the necessary spectral resolution for kinematic modelling but were key in pushing the redshift frontier \citep{Rhoads:2013aa,Oesch:2016aa} beyond cosmic noon \citep[e.g.]{Brammer:2012aa} and in mapping out star formation on spatially resolved scales \citep{Dominguez:2013aa,Nelson:2016wo}. The \textit{JWST} Near Infrared Imager and Slitless Spectrograph (NIRISS) shares the WFSS capability of the NIRCam grism mode, making it another powerful probe of resolved stellar populations \citep{Matharu:2023uj, Pirzkal:2024aa}, but lacks the spectral resolution needed for kinematic analyses.

In order to derive kinematic measurements from grism data, the intrinsic emission line morphology must be deconvolved from distortions caused by velocities and velocity dispersions. This approach has already been successfully applied to 2D spectra from MOSFIRE \citep{Price:2016uv,Price:2020wf}, HST \citep{Outini:2020uz}, and the NIRSpec/MSA \citep{de-Graaff:2024ab}. Some approaches have also been explored for the NIRCam grism data \citep[][]{Nelson:2024aa}, but lack the full instrument forward modelling required to recover kinematic information successfully and reliably. 

At the heart of WFSS data, such as the NIRCam grism data, is the ability to obtain spectra for large samples of galaxies. It is hence crucial to develop an analysis tool that is able to fit objects in a time-frame that allows for large statistical studies. In recent years, the development of optimized gradient-based sampling algorithms has boosted the performance of high-dimensional inference. Furthermore, the use of libraries such as \textsc{jax} \citep{jax2018github}, which enforce differentiable functions and fixed-sized arrays, allows for significant speed increase, specifically with graphics processing units (GPUs). 

 We present here the Grism Emission-line Kinematics tOol (\geko), a Python package which allows for flexible modelling of 2D grism emission line maps through a Bayesian inference framework called \textsc{numpyro} \citep{Phan:2019wc}. We incorporate models for resolved emission line morphologies and kinematics into a full forward modelling of the NIRCam instrument and grism mode. Building on these advances in sampling and inference, we have developed a tool that is state-of-the-art in terms of Bayesian inference in high dimensions and can handle the large parameter spaces needed for fitting complex models. The code is optimized for GPUs, where a galaxy can be fit in approximately twenty minutes, depending on the signal-to-noise of the emission line.  

\geko\ opens the door to various new science cases which focus on building statistical samples of gas kinematics across cosmic time. At Cosmic Noon, the NIRCam grism probes the Paschen lines ($\rm Pa\alpha, Pa\beta$), offering a new probe for dust attenuation when used in harmony with \Ha\ maps from HST \citep[e.g.][]{Liu:2024aa}. Comparing with \Ha\ kinematics \citep[e.g.][]{Wisnioski:2015vx} will offer insight into biases induced by dust attenuation in central regions of galaxies. Furthermore, this instrument will allow us to push the redshift frontier of ionized gas kinematic studies, probing \Ha\ out to $z\approx 6.5$ \citep{Danhaive:2025aa} and \OIII\ out to $z\approx 9$.

This paper is organized as follows. In Sec. \ref{sec:geko-ingredients}, we present the building blocks of \geko\, including the grism forward modelling, and the morphology and kinematics models. In Sec. \ref{sec:mock-tests}, we demonstrate the main recovery tests done on mock data, and in Sec. \ref{sec:real-data} we present an example run of the code on real grism data. In Sec. \ref{sec:discussion}, we discuss some caveats of the code and conclude with future prospects for the development of \geko. \geko\ is
available online at \url{https://github.com/angelicalola-danhaive/geko}, along with scripts to run the mock tests described in this paper, and in-depth tutorials.

\section{\geko\ ingredients} \label{sec:geko-ingredients}

\begin{figure*}
    \centering
    \includegraphics[width=1\linewidth]{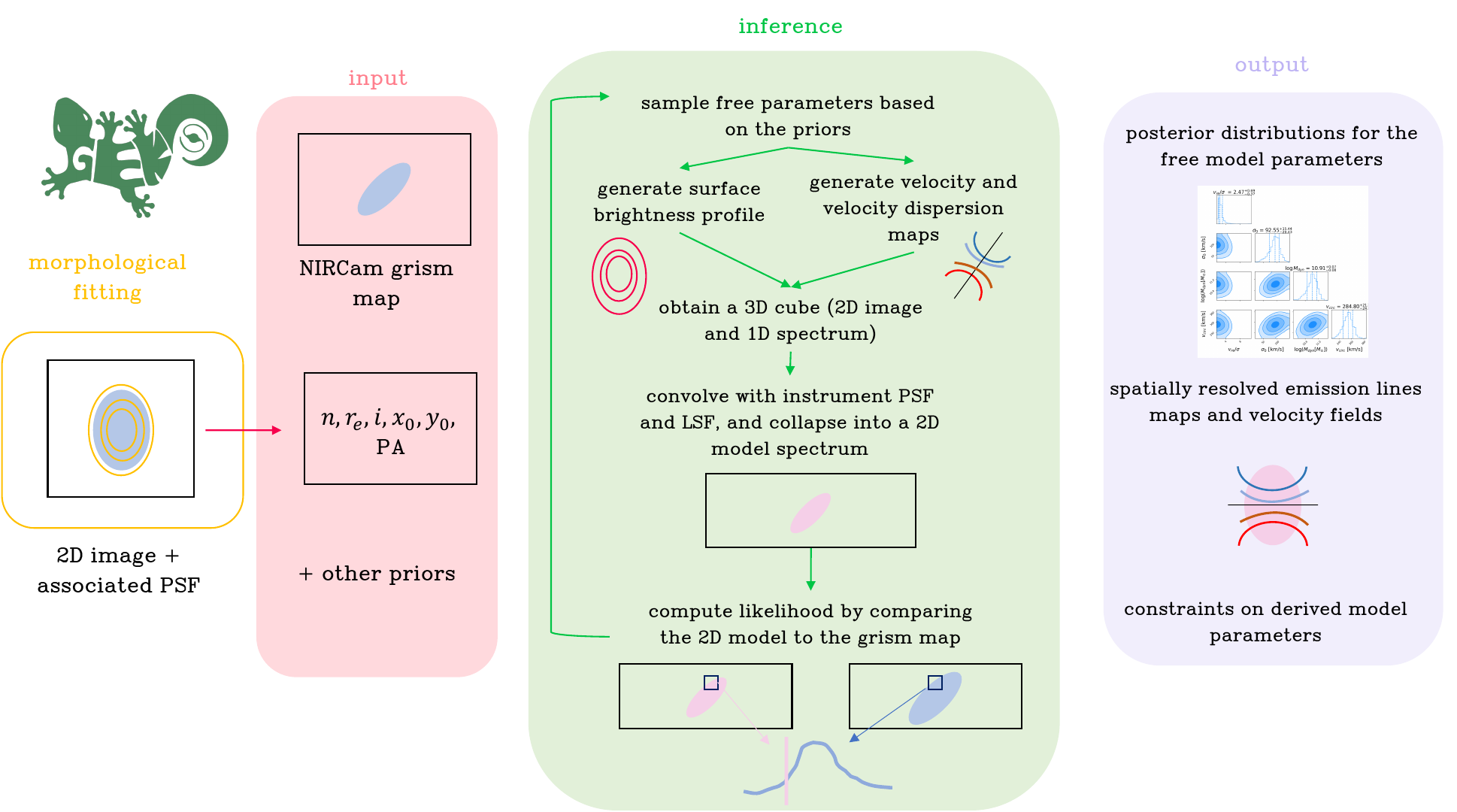}
    \caption{Flowchart of the \geko\ inference process as described in Sec. \ref{sec:geko-ingredients}. The code takes as input continuum subtracted grism data, as well priors for the galaxy's morphology, and then outputs spatially resolved emission line maps and kinematic maps. In the inference process, a mock grism spectrum is constructed based on the current sampled parameters, and is compared to the observed spectrum to evaluate the likelihood.}
    \label{fig:geko-flowchart}
\end{figure*}

\geko\ combines forward-modelling of the grism instrument with a Bayesian inference framework to infer the model parameters of the galaxy. We discuss the currently implemented models in Secs. \ref{sec:morph} and \ref{sec:vel-models}, respectively, and present our approach to forward modelling the grism data (Sec. \ref{sec:forward-mod}). We then detail the inference framework (Sec. \ref{sec:inferece}) and briefly discuss data inputs and post-processing (Secs. \ref{sec:preprocessing} and \ref{sec:postpro}).

It is important to note that \geko\ is a flexible tool, and the specific models and priors chosen will depend on the nature of the data, namely the signal-to-noise of the observations, the size of the galaxy, the redshift, and others. The flexibility allowed in the modelling process naturally needs to reflect the information content of the data in order to not over-fit and be prior dominated. The derived morphological and kinematic parameters will depend on the model chosen and the prior selected for its free parameters. In this section, we present the core morphological (Sec. \ref{sec:morph}) and kinematic (Sec. \ref{sec:vel-models}) models, upon which more complex ones can be built. 

In fact, despite the great benefits of fitting fluxes and velocities in a non-parametric\footnote{We use the term "non-parametric" to describe models, with a high dimensional parameter space, which have no functional form (i.e., pixel-by-pixel fitting).} way, such as modelling clumpy star-formation and merging systems, the degeneracy between kinematics and morphology in the 2D grism maps is strong and hence does not contain enough information for pixel-by-pixel modelling. This effect is even more prominent at high redshift where galaxy sizes are closer to the point-spread function (PSF) resolution. However, when more information is available, such as observations with different position angles or using both grism dispersion directions, then non-parametric modelling can be explored (see Sec. \ref{sec:discussion}). This is particularly relevant for cases where the base models with functional forms are not a good fit for the morphology and/or the kinematics. For example, this is the case for clumpy systems\footnote{Although at $z\sim2$, clumps clearly visible in the flux maps had no strong effect on the ionized gas kinematics \citep[e.g.][]{Genzel:2017aa}.}, merging objects, and clear outflow signatures. To address these more complex objects, there are various additions that could be considered:
\begin{itemize}
    \item[--] a non-parametric grid $\Delta v(x,y)$: add an extra velocity shift to each pixel to account for diversions from the underlying disk velocity field.
    \item[--] a second disk component: this second disk model has a free kinematic centroid so it can be used to model two clumps in one galaxy as well as a merger system where each galaxy is modelled as a disk. In this second case, we also fit for a velocity offset between the two components.
\end{itemize}
These additions will be explored in future version of the code. However, caution must be used when exploring these more complex models, as the data information content needs to match the allowed flexibility. 

Fig. \ref{fig:geko-flowchart} summarizes the \geko\ modelling framework. \geko\ takes a NIRCam grism 2D spectrum of a galaxy as input, along with priors from morphological modelling of the corresponding imaging data. Through a forward modelling inference framework, which accounts for instrumental effects, it then samples the posterior distributions of the model parameters and derives the intrinsic kinematics and spatial distribution of the emission line tracer. It also allows for the derivation of additional parameters such as the rotational velocity at the effective radius, $v_{\text{re}}$, the rotational support, $\rotsupp$, and the dynamical mass $M_{\text{dyn}}$. Within the inference itself, \geko\ generates a mock grism 3D cube of the emission line flux and spectrum, using the chosen morphological and kinematic models, then collapses it onto the detector plane to create a model grism spectrum. This model is compared pixel-by-pixel to the observed grism map to derive the likelihood. We will now discuss each of these steps in more detail.

\subsection{Morphological modelling}\label{sec:morph}
The first step in \geko\ is to model the intrinsic emission line map of the galaxy. As a base model for the surface brightness, we use a \sersic\ profile \citep{Sersic:1968aa}:
\begin{equation}
    I(r) = I_e \exp\Biggl\{ -b_n\left[ \left(\frac{r}{\reff}\right)^{1/n}-1\right]\Biggr\},
\label{eq:\sersic_profile}
\end{equation}
where $I_e$ is the intensity at the half-light radius $\reff$, $n$ is the \sersic index, and $b_n = b(n)$ is approximated by a polynomial\footnote{$b(n) = 2n - 1/3 + 4/405n + 46/25515n^2 +131/1148175n^3 - 2194697/30690717750n^4$.} \citep{Ciotti:1999aa}. This is done because the inverse gamma function typically used to calculate $b_n$ is not differentiable, and hence not \textsc{\textsc{jax}} compatible. The radius $r$ is defined as the distance from the flux centroid on the sky $(x_0,y_0)$, which implies two additional free parameters of the model. This model also needs a position angle, $\rm PA_{\text{morph}}$ and an inclination $i$, or ellipticity $e$:
\begin{align}
    e =& 1 - \frac{b}{a}, \\
    i =& \cos^{-1}\left(\sqrt{\frac{(1-e)^2 - q_0^2}{1-q_0^2}}\right), \label{eq:inc}
\end{align}
where $\frac{b}{a}$ is the axis ratio and $q_0$ is the assumed ratio of scale height to scale length, representing the intrinsic thickness of the disk. This parameter can be set freely within \geko\, but it is chosen as $q_0=0.2$ by default, as our first application of this code \citep{Danhaive:2025aa} is with galaxies at high redshift, where galaxies are assumed to be thick \citep{Wuyts:2016aa, Genzel:2017aa, Price:2020wf,Ubler:2024aa}.  To model the emission line morphology, \geko\ hence fits for seven free parameters, $i$, \PAmorph, $\reff$, $I_e$, $n$, $x_0$, and $y_0$, whose priors are input by the user. Typically, these priors will be derived from modelling of available imaging of the galaxy. The code currently can receive as input modelling results from \textsc{pysersic} \citep{Pasha:2023aa}, which is used to create priors for the morphological parameters (see Sec. \ref{sec:morph-fit}), but the user can also input priors manually for each of the free parameters.

\subsection{Kinematic modelling}\label{sec:vel-models}
We now discuss the base model chosen for the spatially resolved kinematics of the emission line, which is used to generate the velocity and velocity dispersion grids needed to disperse an object onto the grism plane. The simplest model for a galaxy is an idealized thin disk model. Because these exponential disks are embedded in dark matter haloes, their kinematics are well modelled by the arctangent function \citep{Courteau:1997uw, Miller:2011aa}:
\begin{equation}
    v_{\text{rot}}(r_{\text{int}}, r_{\text{t}}, V_a) = \frac{2}{\pi}V_a\arctan{\frac{r_{\text{int}}}{r_{\text{t}}}},
\end{equation}
where $v_{\text{rot}}$ is the rotational velocity at a given radius $r_{\text{int}}$ in the intrinsic galaxy plane, $V_a$ is the asymptotic value that the arctangent rotation curve tends to at large radii $r_{\text{int}} \rightarrow \infty$ , and $r_{\text{t}}$ is the turn-around radius of the rotation curve.
To project this velocity on the observation plane, we need to account for the galaxy's inclination $i$:
\begin{equation}
        v_{\text{obs}}(x,y) = v_{\text{rot}}(R_{\text{int}}, R_{\text{t}}, V_a)\cdot\sin{i}\cdot\cos{\phi_{\text{int}}},
        \label{eq:vobs}
\end{equation}
where $\phi_{\text{int}}$ is the polar angle coordinate in the galaxy plane. The intrinsic galaxy coordinates are centred on a velocity centroid\footnote{The user can choose to leave this as a free parameter or fix it to the centre of the surface brightness profile $(x_0,y_0)$}. For the full derivation of $v_{\text{obs}}$, see Appendix \ref{sec:annexe_vel}. We fix the inclination correction $i$ to the velocity curves to the morphological inclination (Eq. \ref{eq:inc}).

We model a constant isotropic velocity dispersion $\sigma_0$ across the disk based on deep adaptive optics imaging spectroscopy studies at lower redshift \citep{Genzel:2008ug, Forster-Schreiber:2018aa}:
\begin{equation}
    \sigma(x,y) =  \sigma_0.
\end{equation}
Although some high-resolution observations of molecular and atomic gas in the local Universe have shown radially declining velocity dispersions, the observed radial changes are well below the NIRCam grism spectral resolution \citep[for a more detailed discussion see][and references therein]{Ubler:2019vg}. We also introduce a velocity offset $v_0$, which acts as an effective redshift offset in the observed wavelength of \Ha\ by shifting the full 2D emission line map.
The free kinematic parameters are hence the inclination $i$, the position angle $\text{PA}$, the asymptotic velocity $V_a$, the turn-around radius $r_t$, the intrinsic velocity dispersion $ \sigma_0$, the velocity centroid $(x_{v},y_{v})$, and the velocity offset $v_0$.

\subsection{Forward-modelling the instrument}\label{sec:forward-mod}
\renewcommand{\arraystretch}{2}
\begin{table*}
    \centering
    \begin{tabular}{c|c|c|p{6cm}}
    \textit{Name} & \textit{Parameter} \textit{Description} & \textit{Prior} & \textit{Prior Description} \\ \hline 
$\text{PA}_{\text{morph}}$ & Position angle of \Ha\ morphology & Normal$(\mu_{\rm PA}, \sigma_{\rm PA})$ & Gaussian prior, in degrees. \\ \hline
$i$ & Inclination angle & TruncNormal$_{[0, 90]}(\mu_i, \sigma_i)$ & Truncated Gaussian prior. \\ \hline
$A$ & Amplitude of brightness & TruncNormal$_{[0, \infty]}(\mu_A, \sigma_A)$ & Truncated Gaussian prior, in mJy, based on the integrated continuum subtracted 1D spectrum. \\ \hline
$r_\text{e}$ & Effective radius & TruncNormal$_{[r_\text{min}, r_\text{max}]}(\mu_r,\sigma_r)$ & Truncated normal prior in pixels. \\ \hline
$n$ & \sersic\ index & Normal$(\mu_n, \sigma_n)$ & Gaussian prior. \\ \hline
$x_0$ & X--coordinate of image centre & Normal$(\mu_{x_0}, \sigma_{x_0})$ & Gaussian prior. \\ \hline
$y_0$ & Y--coordinate of image centre & Normal$(\mu_{y_0},\sigma_{y_0})$ & Gaussian prior, in pixels. \\ \hline \hline

$\text{PA}_{\text{kin}}$ & Position angle of kinematics & Normal$(\mu_{\rm PA}, \sigma_{\rm PA})$ & Same prior as $\text{PA}_{\text{morph}}$, but fit independently. \\ \hline
$V_a$ & Asymptotic rotational velocity & Uniform$(-1000, 1000)$ & Uniform prior from -1000 to 1000 km/s by default. \\ \hline
$\sigma_0$ & Intrinsic velocity dispersion & Uniform$(0, 500)$ & Uniform prior from 0 to 500 km/s by default. \\ \hline
$r_t$ & Turnover radius & Uniform$(0, r_\text{e})$ & Uniform prior from 0 to $r_\text{e}$ pixels. \\ \hline
$v_0$ & Systemic velocity & Normal$(0, 50)$ & Gaussian prior with mean 0 km/s and standard deviation 50 km/s. \\ \hline
$x_v$ & X--coordinate of kinematic centre & Normal$(\mu_{x_0}, \sigma_{x_0})$ & Gaussian prior, in pixels. \\ \hline
$y_v$ & Y--coordinate of kinematic centre & Normal$(\mu_{y_0}, \sigma_{y_0})$ & Gaussian prior, in pixels. \\ 
    \end{tabular}
    \caption{Description of morphological (top section) and kinematic parameters (bottom section) being fit as free parameters in our \geko\ modelling, along with their default priors. For all of the parameters $p$ with Gaussian priors based on morphological information, we refer to the chosen mean as $\mu_p$ and its uncertainty $\sigma_p$. The values chosen for these priors are all flexible to be modified by the user.}
    \label{tab:priors}
\end{table*}

Using the emission line morphology (Sec. \ref{sec:morph}) and kinematic (Sec. \ref{sec:vel-models}) models, we can build a 3D model cube $\text{I}(x, y, \lambda)$ which has the emission line flux in each pixel on the 2D plane, and then contains the kinematics encoded in the third spectral axis. The spectrum in each pixel is defined as a Gaussian centred on the observed wavelength of the chosen emission line $\lambda_{\rm obs} = (1+z)\lambda_{\rm el}$, and then shifted and broadened by the velocity and velocity dispersion in that pixel, respectively. In this first version of the code we only model one emission line at a time and assume no continuum emission. In fact, the grism data is continuum-subtracted using median filtering (Sec. \ref{sec:preprocessing}) prior to the modelling. This allows us to directly model the intrinsic morphology and kinematics of a specific line.

To derive a model grism map from $\text{I}(x, y, \lambda)$, we need to model the instrument itself and map each pixel from the 3D model plane to the 2D grism plane, which we detail in the following sections.

\subsubsection{The grism dispersion function}
The key function needed to model the grism instrument is the grism dispersion function $\Delta x(x_0,y_0,\lambda)$, which projects a pixel the image plane $(x_0,y_0)$, emitting at a wavelength $\lambda$, to a position on the grism detector plane $(x_\lambda,y_\lambda)$ . This function returns the pixel offset
\begin{equation}
    \Delta x = x_\lambda - x_0,
\end{equation}
i.e. the difference between where the pixel is dispersed to on the grism plane and the initial pixel position. The dispersion function can be written as a polynomial
\begin{equation}
    \Delta x(x_0,y_0,\lambda) = \alpha_1 + \alpha_2\cdot\lambda + \alpha_3\cdot\lambda^2 + \alpha_4\cdot\lambda^3, 
    \label{eq:disp}
\end{equation}
where the coefficients $\alpha_i$ depend on $x_0,x_0^2,y_0$ and $y_0^2$. The best fit parameters for the constant factors in this function have been calibrated during commissioning for each module/filter/pupil combination \citep[for full details, see][]{Sun:2023ab}. These commissioning observations were obtained through Program \#1076 (PI: Pirzkal) during
the commissioning phase of the instrument \citep{Rieke:2023va}. However, when fitting combined spectra from multiple exposures, the $x_0,y_0$ information is lost as in every exposure, the source is in a different position on the detector. We therefore set $x_0,y_0$ to be at the centre of the detector. The $\alpha$ coefficients are expressed as polynomials which depend on $x_0$ and $y_0$ in the following way:

\begin{align*}
    \alpha_1 =& \alpha_{10} + \alpha_{11}\cdot x_0 + \alpha_{12}\cdot y_0 + \alpha_{13}\cdot x_0^2 + \alpha_{14}\cdot y_0^2 + \alpha_{15}\cdot x_0y_0 \\
    \alpha_2 =& \alpha_{20} + \alpha_{21}\cdot x_0 + \alpha_{22}\cdot y_0 + \alpha_{23}\cdot x_0^2 + \alpha_{24}\cdot y_0^2 + \alpha_{25}\cdot x_0y_0\\ 
    \alpha_3 =& \alpha_{30} + \alpha_{31}\cdot x_0 + \alpha_{32}\cdot y_0 \\
    \alpha_4 =& \alpha_{4}
\end{align*}

The $0^{\rm th}$ order factors ($\alpha_{i0}$) are of the order of $\mathcal{O}\sim 10^{3}-10^{0}$, whereas the higher order factors ($\alpha_{ij=1,2,3,4,5}$) are much smaller ($\mathcal{O}\sim 10^{-3}-10^{-7}$). Therefore, setting  $x_0,y_0$ to be at the centre of the detector has a relatively negligible effect effect. Fitting exposure per exposure is possible in the \geko\ framework but would require some manual adaptations made by the user.
We also set
\begin{equation}
    y_\lambda = y_0 \longrightarrow \Delta y = 0,
\end{equation}
meaning we assume no spectral tracing. This assumption is based on the fact that we model a single emission line, and spectral tracing has a negligible effect on the order of the few pixels covered by the line. In this simple framework, we use data-specific variables, such as filter, pupil, module, and wavelength space, which allow for the dispersion to be accurately modelled for each specific observation.

\subsubsection{The detector space}\label{sec:model_space}

In practice, the dispersion from image space to grism space can be done in any resolution by oversampling the spatial and spectral axes of the detector space when moving to the modelling space. This allow us to disperse our model image in high resolution, improving accuracy before sampling back down to the detector space, where the comparison between the model and the observations takes place. The amount of oversampling can be tuned to jointly optimize speed and accuracy. The default values are a factor of 5 oversampling for the spatial axes and 9 for the spectral axis \footnote{Oversampling by an uneven number keeps the symmetry of the arrays.}.

\geko\ fits for intrinsic model parameters, which are corrected for PSF and line-spread-function (LSF) effects. Hence, when projecting our model on the grism detector plane, we need to first account for the LSF broadening. 
To do so, each pixel's effective dispersion is computed as the square sum of the intrinsic velocity dispersion $\sigma_{0}$ and the LSF $\sigma_{\text{LSF}}$
\begin{equation}
    \sigma_{\text{eff}} = \sqrt{\sigma_{0}^2 +  \sigma_{v,\text{LSF}}^2}.
\end{equation}
where $\sigma_{v,\text{LSF}}$ is the LSF computed in velocity space,
\begin{equation}
    \sigma_{v,\text{LSF}} = \frac{\sigma_{\text{LSF}}}{\lambda/c},
\end{equation}
where the speed of light $c$ is in km/s. 

We also account for the filter-specific PSF. In \citet{Danhaive:2025aa}, the PSF is approximated using the model PSFs (mPSFs) from \citet{Ji:2024aa} constructed by mosaicking WebbPSF models repeatedly over the field identically to our exposure mosaics and then measuring the average PSF. However, the user gives their chosen PSF as part of the \geko\ inputs. We convolve our model cube with this 2D PSF kernel and the 1D LSF kernel:
\begin{equation}
    \text{I}(x, y, \lambda) \otimes \text{PSF}(x, y) \otimes \text{LSF}(\lambda).
    \label{eq:convolution}
\end{equation}
The final grism spectrum is then computed by projecting each pixel of the convolved 3D cube onto a detector row using the dispersion function (Eq. \ref{eq:disp}), then summing over each row. For one row $k$ of the grism array $G$, the grism model is computed as
\begin{equation}
 \mathbf{G}_{i,k} = \Sigma_j \mathbf{I}_{i,j} \cdot \mathbf{C}_{j,k} 
\end{equation}
where $C$ is contains the normalized expansion of each pixel in the dispersion space, based on the velocity and velocity dispersion in that pixel. This framework allows us to self-consistently account for the PSF and the LSF of the instrument and recover intrinsic values that are PSF/LSF corrected.

\subsection{Inference and sampling}\label{sec:inferece}

The grism instrument forward model is integrated in a Bayesian inference framework that allows us to obtain the posterior distributions of the free model parameters by fitting the observed grism data. To achieve this, we use \textsc{numpyro}, as described in the following section.

\subsubsection{\textsc{numpyro} and \textsc{jax}}

\textsc{numpyro} \citep{Bingham:2018tw,Phan:2019wc} is a probabilistic programming language that provides a Numpy back-end to Pyro. It is based on \textsc{jax} \citep{jax2018github}, which is a library allowing for program transformations, such as automatic differentiation and just-in-time (JIT) compilation, of Python functions. Automatic differentiation \citep{Griewank:2000} describes a set of computational methods for the rapid, automated calculation of mathematical derivatives. They are highly efficient when using Markov Chain Monte Carlo (MCMC) algorithms, such as Hamiltonian Monte Carlo \citep[HMC; ][]{Neal:2011aa}, which rely on the computation of gradients. JIT compilation works on two planes. First, it compiles the program initially and translates it to machine language in order for it to be executable. Then, it continues to work in parallel with the execution of the program (hence the name) to spot frequently repeated sections of the code and compiles them to further speed up the run time. This boosts the performance of \textsc{numpyro} compared to otherwise similar inference tools.

In order to sample the posterior distributions of our free model parameters, we use The No-U-Turn-Sampler \citep[NUTS; ][]{Hoffman:2011wm}, which is an adaptation of an HMC algorithm. HMC uses the theory of Hamiltonian mechanics to suppress the random walk behaviour of other MCMC methods, treating the walkers as particles moving in the potential dictated by the negative joint density of the parameters that are being sampled. However, HMC requires the user to set values for the step size $\epsilon$ and the number of steps $L$ (taken by the algorithm before generating a proposal), a choice which heavily impacts the performance of the algorithm. NUTS is an adaptation of HMC where the number of steps $L$ and the step size $\epsilon$ are automatically computed, removing the task of having to hand-tune these parameters. The \textsc{numpyro} implementation of the NUTS sampling also allows the user to hand-tune the adaptive path length and the mass matrix. For all of the reasons detailed above, this is the algorithm we chose for our sampling. We have also implemented the nested sampling approach from \textsc{numpyro}, which the user can select.

This implementation is optimized for GPUs, even though it can also be run on CPUs. As an example, fitting one chain (or multiple if parallelized) for 500 warm-ups and 1000 samples takes $~2-5$h using local CPUs, but $\sim 10-30$ mins on a cluster GPU. This allows \geko\ to fit large samples in a relatively short time.

\subsubsection{Physically motivated priors}\label{sec:priors}
We will now describe the priors set within \geko\ for the free model parameters, which are also summarized in Tab. \ref{tab:priors}.

In order to alleviate the kinematics-morphology degeneracy, we need to impose some priors on the morphological parameters using the available imaging data. Morphological fitting of imaging is not done within \geko, which instead takes an input catalogue of modelling results and uncertainties. For each of the morphological free parameters, $i$, \PAmorph, $\reff$, $I_e$, $n$, $x_0$, and $y_0$, \geko\ defaults to Gaussian priors, centred on the best-fit value and with uncertainties typically doubled to account for differences in morphology between the continuum probed in the imaging data and the emission line itself. In addition, we include some physical bounds. The flux $I_e$ and the inclination $i$, cannot be negative, so we truncate their priors at zero.

For the kinematic parameters, we first fix the velocity centroid and the inclination to their morphological counterparts. The kinematic position angle is free to differ from the morphological one, but both have the same prior. The velocity offset $v_0$, which effectively accounts for any offset in the redshift, has a Gaussian prior centred on zero. The rest of the free parameters, the asymptotic velocity $V_a$, the turn-around radius $r_t$, and the intrinsic velocity dispersion $\sigma_0$, have uniform priors. Both $r_t$ and $\sigma_0$ are have priors truncated at zero, but $V_a$ can be both positive or negative to allow for flexibility on which portion of the galaxy is blue/red shifted.

\subsubsection{Likelihood evaluation}
As \geko\ samples the prior distributions within the \textsc{numpyro} framework, it evaluates the likelihood function to inform the next sample and construct the posterior distribution. The likelihood function is Gaussian, and for a set of parameters $\vec{p}$, is defined as
\begin{equation}
    \ln \mathcal{L}(D|\vec{p}) = - 0.5 \cdot \sum^{N}_{i}  \left(\frac{(D_i - G_i(\vec{p}))^2}{\sigma_i} + \ln (2\pi\sigma_i^2)\right),
\end{equation}
where $D_i$ is the flux in the $i^{\rm th}$ pixel of the observed 2D grism data $D$, $\sigma_i$ is its uncertainty, $G_i$ is the flux in the modelled grism spectrum, and $N$ is the total number of pixels in the grism map. The full grism spectrum is typically cropped around the chosen emission line to reduce the number $N$, since most pixels will not be informative in the fitting process.

\subsection{\geko\ inputs}\label{sec:preprocessing}

There are two main inputs that \geko\ requires, as shown on Fig. \ref{fig:geko-flowchart}. The first is a continuum subtracted grism map centred on the emission line chosen by the user. In order to obtain continuum-subtracted emission-line maps in \citet{Danhaive:2025aa}, we use row-by-row median filtering \citep{Kashino:2023wv} which removes continuum emission from the object itself and any neighbouring objects that may leave traces in the spectrum. This 2-step (iterative) median filtering technique aims to ensure that the continuum is accurately removed but the emission line flux is not harmed. We then cut the grism 2D spectra symmetrically around the emission line so that the region being fitted is not too large. This is implemented within \geko, within the preprocessing module, but can also be done by the user prior to fitting. 
The second input consists of priors for the morphology of the galaxy. As described above (Sec. \ref{sec:priors}), these priors are chosen by the user, typically based on prior modelling of the image in a relevant band. The uncertainties on these priors can be chosen based on the band used. We plan on incorporating morphological fitting, prior to the \geko\ modelling, within \geko\ in a future release. Other inputs include redshift and rest-frame wavelength of emission line, or simply the observed wavelength of the emission line.

\subsection{Post-processing}\label{sec:postpro}

From the inference we obtain posterior distributions, and hence best-fit values and uncertainties, for all of the free parameters in the model. From these, we can also derive posteriors for key physical parameters such as the rotational velocity at the effective radius $v_{\text{rot}}(r_{\rm e})$, the rotational support $\rotsupp$, the circular velocity $v_{\text{circ}}$, and the dynamical mass $M_{\text{dyn}}$. The rotational velocity is simply obtained from the rotation curve, corrected for inclination:
\begin{equation}
    v_{\text{rot}}(r_{\rm e}) = v_{\text{obs}}(r_{\rm e})/\sin i,
\end{equation}
where $v_{\text{obs}}$ is defined by Eq. \ref{eq:vobs}.

Assuming virial equilibrium, the circular velocity of galaxies is computed by combining the effects of gravity and turbulence-induced pressure. The first term is the rotational velocity, and the latter is an asymmetric drift correction to account for the pressure support \citep{Burkert:2010aa,Newman:2013aa, Wuyts:2016aa}. For an exponential disk, it takes the form:

\begin{equation}
    v_{\text{circ}}(r) = \sqrt{v_{\text{rot}}^2(r) + 2(r/r_{\rm s})\sigma_0^2}
\end{equation}
where $r_{\rm s}$ is the disk scale length. At the effective radius $r_{\text{e}}$ where we compute the circular velocity  $v_{\text{circ}}(r_{\text{e}})$, $2(r_{\text{e}}/r_{\rm s}) = 3.35$. We compute the total dynamical mass following \cite{Price:2020wf}:
\begin{equation}
    M_{\text{dyn}} = k_{\text{tot}}\frac{r_{\text{e}}v^2_{\text{circ}(r_{\text{e}})}}{G},
\end{equation}
where $G$ is the gravitational constant and $k_{\text{tot}}$ is the virial coefficient. Because we have modelled our galaxies with $q_0=0.2$, in \citet{Danhaive:2025aa} we choose $k_{\text{tot}}=1.8$ as it is the coefficient for galaxies with $q_0=0.2$ and $n\sim 1-4$ \citep{Price:2022aa}. We note that the users are able to directly compute new posteriors based on the main parameter posteriors, allowing them to modify the choice of certain parameters such as $k_{\text{tot}}$. In the remaining of the paper, we define the rotational velocity at $\reff$ as $v_{\text{rot}}(r_{\rm e}) \equiv \vre$ for simplicity.

\section{Tests with Mock Data} \label{sec:mock-tests}
\begin{figure}
    \centering
    \includegraphics[width=0.7\linewidth]{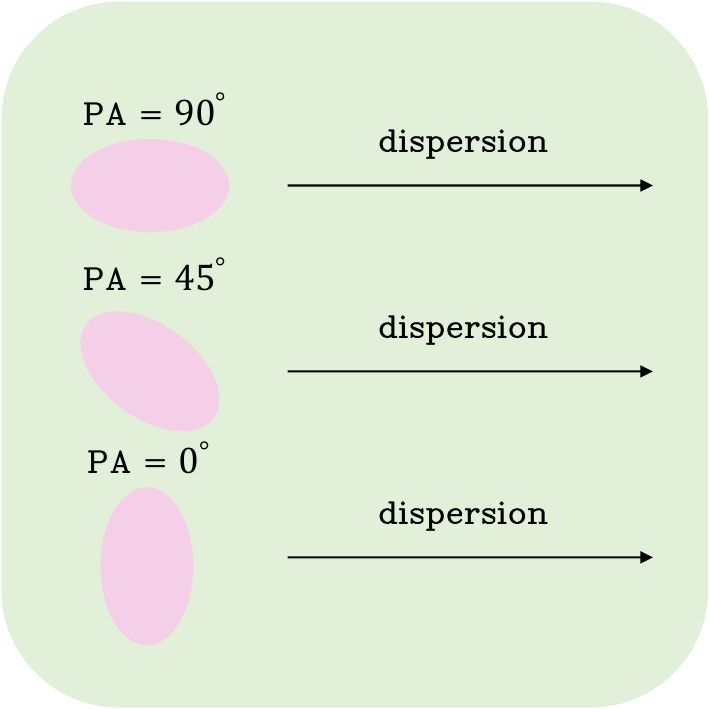}
    \caption{Schematic illustration of the position angles ($\rm PA$) as defined within the \geko\ framework. The PA is measured as the angle between the negative x-axis and the galaxy's major axis, as shown in the $\rm PA = 45^{\circ}$ case.}
    \label{fig:pa-cartoon}
\end{figure}

\begin{figure}
    \centering
    \includegraphics[width=1\linewidth]{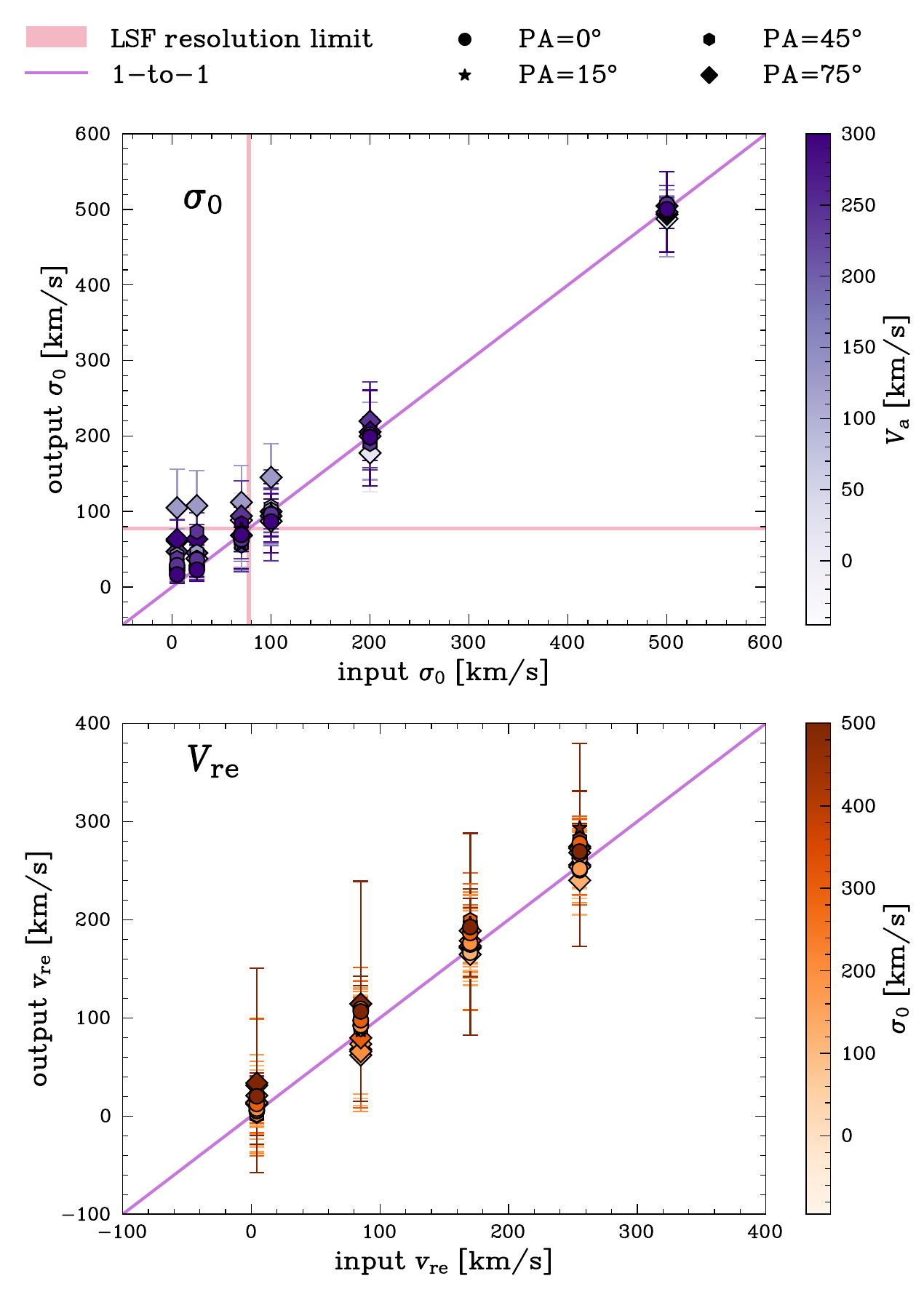}
    \caption{Test results for the recovery of kinematics, using noisy mock grism data, for a range of position angles, $\rm PA = 0^{\circ},15^{\circ},45^{\circ},75^{\circ}$, asymptotic velocities $V_a = 5,100,200,300$ km/s, and velocity dispersions $\sigma_0 = 5,25,70,100,200,500$ km/s. For clarity, we do not show the $\rm PA = 90^{\circ}$ case, but we show the average results on Fig. \ref{fig:real_test_param_mean_curves}.}
    \label{fig:real_test_param}
\end{figure}

In this section, we will show the capabilities of \geko\ in fitting mock data, which is generated using the same forward modelling framework as detailed above (Secs. \ref{sec:morph}, \ref{sec:vel-models}, and \ref{sec:forward-mod}). The goal is to quantify the recovery and the level of uncertainty obtained for different model parameters in different scenarios, in order to best understand the capabilities and caveats of the code. Also, it allows us to show and discuss the main outputs of the code. We run our tests on mock galaxies observed in the F444W filter (both in imaging and grism), emitting \Ha\ at $\lambda_{\text{obs}} = 4.5 \mu$m, corresponding to a redshift of $z\approx 5.8$. At $\lambda_{\rm obs}$, the spectral resolution is $\sigma_{\rm LSF} \approx 75$ km/s. In the tests, we vary observational effects such as position angle on the sky, inclination, and S/N of the grism data, as well as kinematic (rotation and velocity dispersion) and morphological (size and \sersic\ index) parameters.

It is important to note that while we try to explore a range of different configurations, each test result is dependent on the choices made. For example, high S/N will inevitably decrease uncertainties, and higher velocities will be easier to constrain than low ones (due to the pixelisation). When fixing certain parameters that are not varying in a given test, we try to choose values that fall in the middle of the range, so not at the worst or best end in terms of accuracy of recovery. The same caveat applies to the choice of filter and observed wavelength. The spectral resolution of the grism decreases with wavelength, and so the results of the testing would vary accordingly. The python script to run mock tests can be found in the Github repository\footnote{\url{https://github.com/angelicalola-danhaive/geko}}. We encourage users to perform appropriate tests when using \geko.

\subsection{Constructing mock observations}

We begin by simulating the galaxy on the sky plane using a \sersic\ profile 
with $I(\reff) = 1$, and set the true value of $\reff$ as
\begin{equation}
    \reff = \frac{1.676}{0.4}r_t,
    \label{eq:re-rt}
\end{equation}
so that the input value of $r_t$ is intrinsically linked to the input value of $\reff$. However, in the fitting Eq. \ref{eq:re-rt} is not imposed, and both $\reff$ and $r_{\rm t}$ are free parameters of the model (Sec. \ref{sec:priors}). We only use Eq. \ref{eq:re-rt} to generate a physically motivated mock galaxy \citep{Miller:2011aa}. The ellipticity and position angle are set to their chosen kinematic counterparts. For the testing, we set $i = 60^{\circ}$. Based on the input kinematic parameters, we also generate mock velocity and velocity dispersion maps, to create a mock 3D cube.

To construct the mock grism spectra, we use our forward modelling framework to go from the direct image space to the grism space (see Sec. \ref{sec:model_space}). We convolve the intrinsic\footnote{no PSF-convolution or artificial noise} 3D cube with the PSF and LSF (Eq. \ref{eq:convolution}), then project it onto the grism detector space. We add noise to the final 2D spectrum with $\sigma_{\text{noise,grism}} = \beta_{\text{grism}}/\text{SN}_{\text{max,grism}}$ where $\beta$ is the value of the highest flux pixel in the grism image. This allows us the produce a noisy mock grism image, and then test the ability of the code to recover the true morphological and kinematic parameters used to generate it. We note that the $\text{SN}_{\text{max,grism}}$ is different from the integrated S/N, which we compute from the final generated image.

\subsection{Idealized tests}\label{sec:ideal-test}

In order to study the convergence and any intrinsic biases of this model, we first run tests on idealized data. We do not include the PSF and LSF convolutions in the grism dispersions. We also do not add noise to the grism data. An important parameter when running all of these tests is the position angle of the galaxy with respect to the grism dispersion direction. We set the dispersion direction to $\rm PA = 90^{\circ}$, meaning that the position angle is defined with respect to the positive y-axis. On Fig. \ref{fig:pa-cartoon}, we show schematically how the position angle is defined with respect to the grism dispersion direction. If the galaxy is oriented parallel to the dispersion direction ($\rm PA = 90^{\circ}$), then the degeneracies between the morphology and kinematics are strongest, and the effects of the velocity dispersion are closely mimicked by those of velocity gradients from the rotation. On the other hand, the $\rm PA = 0^{\circ}$ case provides the best kinematic constraints. Figs. \ref{fig:PA0-deg} and \ref{fig:PA90-deg} show the joint posterior distributions of $\disp$ and $V_a$, for the $\rm PA=0^{\circ}$ and $\rm PA=90^{\circ}$ cases respectively, and highlight the degeneracy in the $PA=90^{\circ}$ case which disappears when $\rm PA=0^{\circ}$.

The results of these tests are summarized in Appendix \ref{app:ideal-tests}. We note a few important highlights. First, as shown on Fig. \ref{fig:ideal_test_param}, the recovery of $\disp$ worsens as it drops below the $\disp\approx 80$ km/s. This is due to the pixelisation of the grism data, where each pixel in the detector spans $\Delta\lambda = 0.001~\mu \rm m$. Even in these idealised tests where we do not convolve with the LSF, we are still limited by the intrinsic detector pixel resolution. Hence, at values $\disp\lesssim 80$ km/s, the dispersion is over-estimated. However, we note that, especially for low PAs, meaning that the galaxy is close to perpendicular to the dispersion direction, the recovery is satisfactory down to $\disp\approx 25$ km/s, and only really breaks down at smaller $\disp$ values. This suggest that in specific cases, low velocity dispersions can be recovered, albeit with relatively large uncertainties. Having high velocities helps with this recovery because it spreads out the emission line, and offers more pixels where it can be measured. The recovery of $v_{\rm re}$ is good for all values of $\disp$ and  $v_{\rm re}$, with uncertainties increasing with increasing PA and $\disp$. 

\subsection{Realistic tests}\label{sec:real-tests}

We now reproduce realistic mock observations, where the grism data is convolved with the LSF and PSF of the instrument, and contains artificially added Gaussian noise, and study how well we are able to constrain the kinematic properties of the galaxies. We summarize the tests done and which parameters are varied or kept constant on Tab. \ref{tab:tests}. 

\subsubsection{Kinematics test}\label{sec:kin-test}
\begin{figure}
    \centering
    \includegraphics[width=1\linewidth]{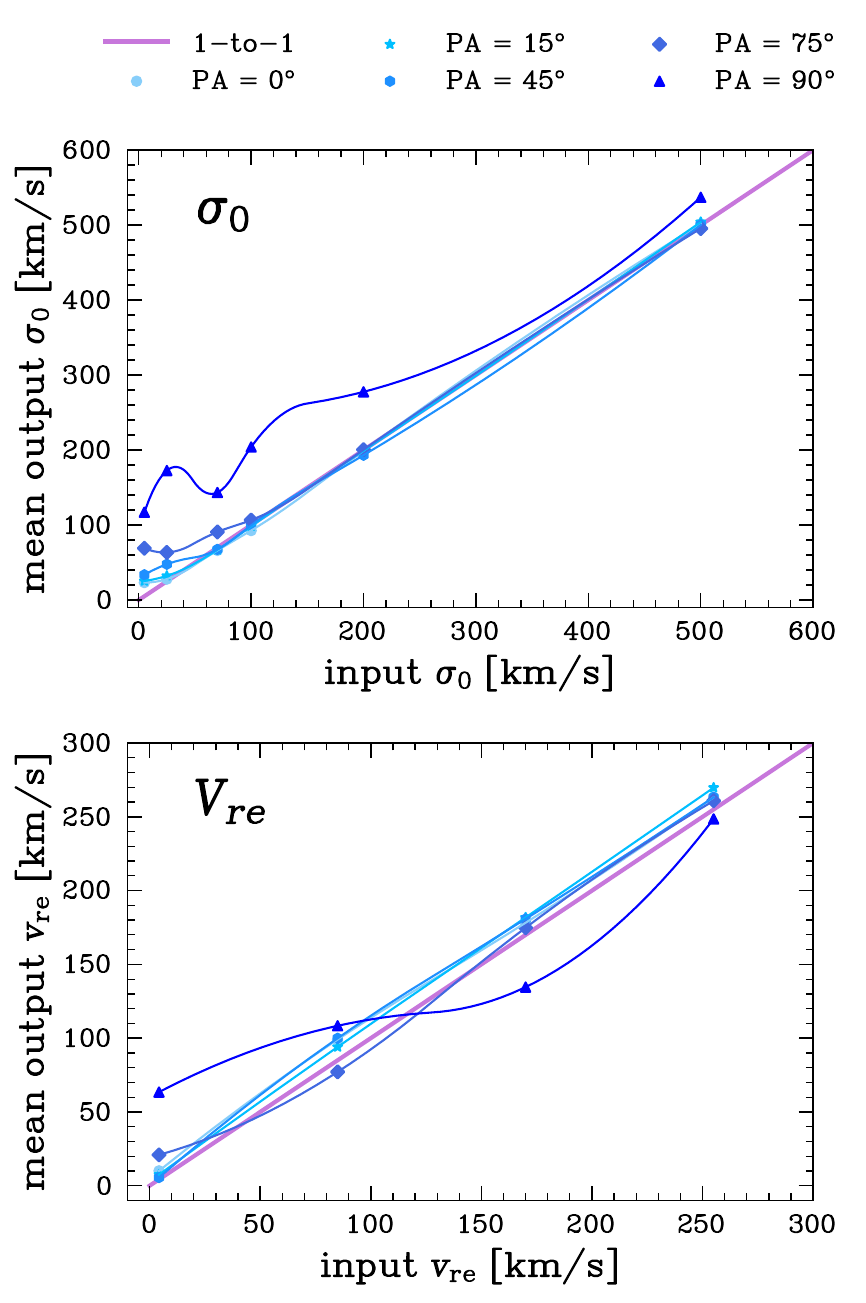}
    \caption{Mean inferred values for $\sigma_0$ and $v_{\rm re}$, for the corresponding input true values, for each position angle $\rm PA = 0-90^{\circ}$. The means are taken over all of the $V_a$-$\sigma_0$ combinations tested and shown on Fig. \ref{fig:real_test_param}. When the galaxy is parallel to the dispersion axis, $\rm PA = 90^{\circ}$, the kinematics cannot be well constrained due to the strong degeneracies.}
    \label{fig:real_test_param_mean_curves}
\end{figure}

\begin{figure}
    \centering
    \includegraphics[width=1\linewidth]{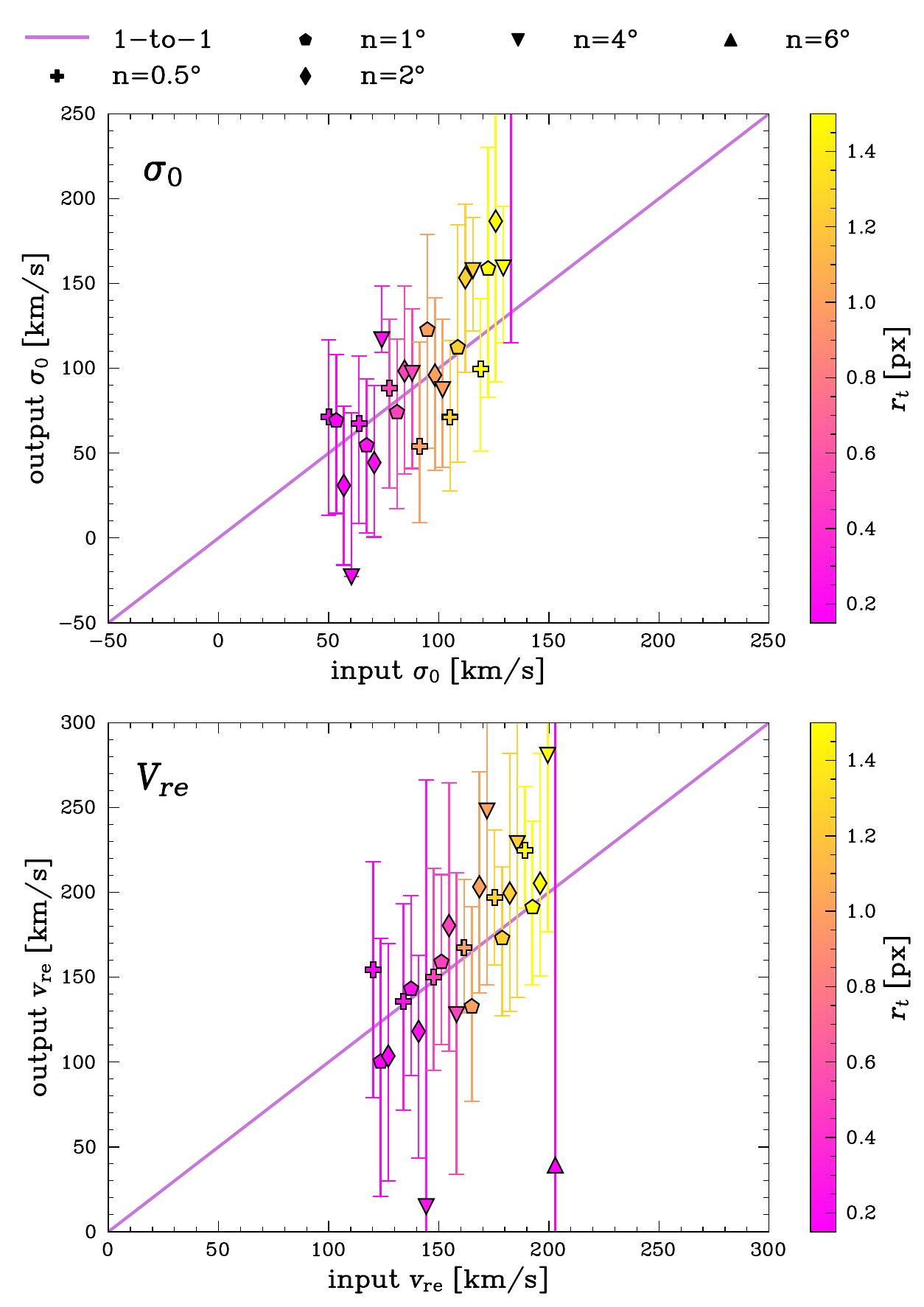}
    \caption{Test results for the recovery of kinematics, using noisy mock grism data, for a range of turn-around around radii $r_{\rm t} = 0.25, 0.5, 1,1.25,1.5$ px (which sets the effective radii $r_{\rm e} \approx 1, 2, 4,5,6 \rm \thinspace px \approx 0.06,0.13,0.25, 0.38''$ following Eq. \ref{eq:re-rt}) and \sersic\ indices $n=0.5,1,2,4,6,8$. The position angle of the mock galaxies is set to $\rm PA = 45^{\circ}$, and the kinematics to $V_a =200$ km/s and $\sigma_0 = 100$ km/s. The input kinematic values are fixed, but we spread them out on the plot artificially, for clarity purposes.}
    \label{fig:size-test}
\end{figure}

\begin{figure}
    \centering
    \includegraphics[width=1\linewidth]{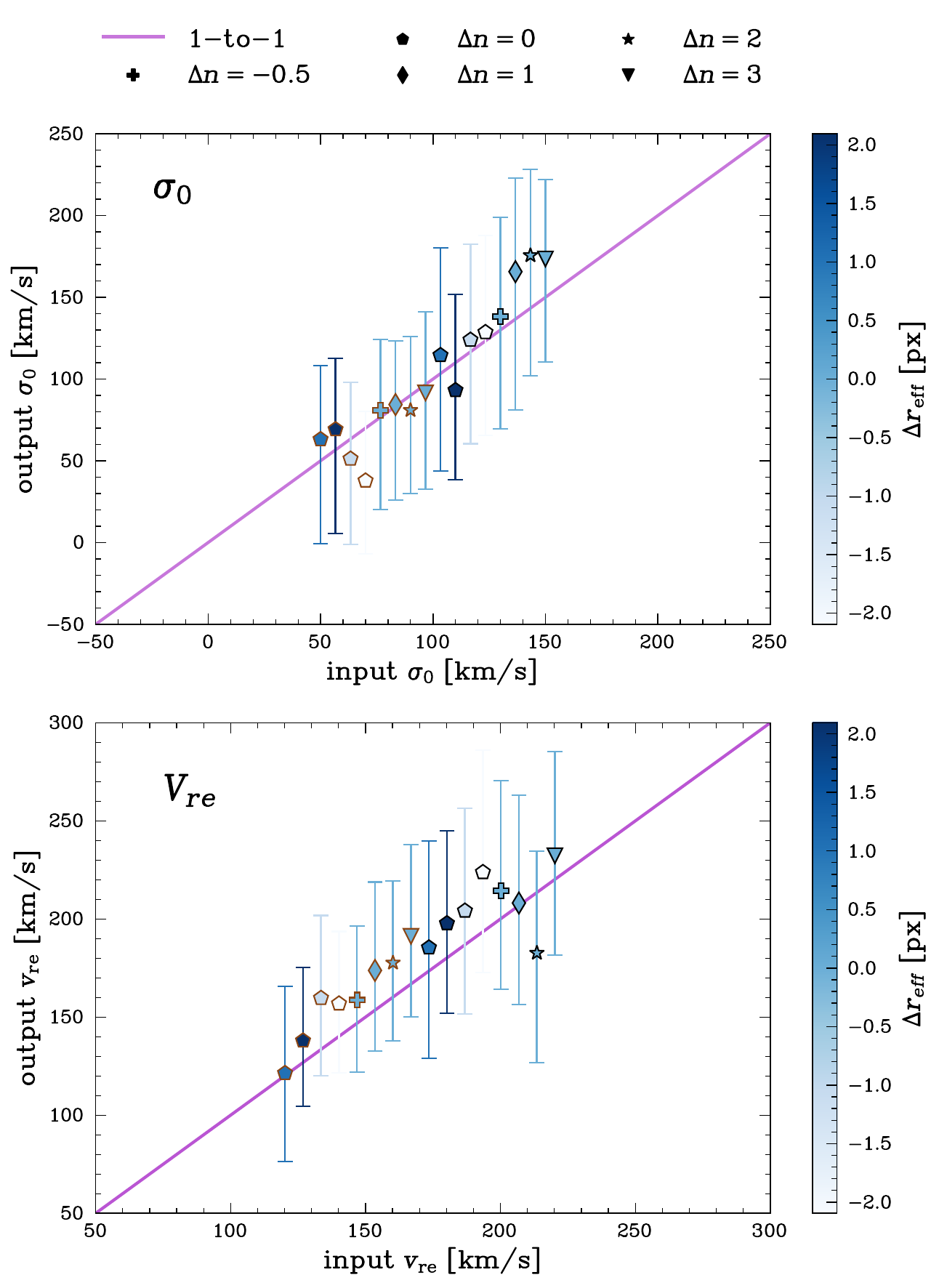}
    \caption{Test results for the recovery of kinematics, using noisy mock grism data, when the mean of the prior for the effective radius $\reff$ is offset from the truth by  $\Delta r_{\rm e}\approx -2,-1,1,2$ px, and the mean of the prior for the \sersic\ index by $\Delta n = -0.5, 1,2,3$. The true values are set to $r_{\rm e} \approx 4$ and $n=1$. When one parameter is varying, the other is fixed to their true value. The position angle of the mock galaxies is set to $\rm PA = 45^{\circ}$, and the kinematics to $V_a =200$ km/s and $\sigma_0 = 100$ km/s. The test is done for mock grism data with with $\text{SN}_{\text{max,grism}} = 10$ (black outlines) and $\text{SN}_{\text{max,grism}} = 20$ (brown outlines). The input kinematic values are fixed, but we spread them out artificially for clarity.}
    \label{fig:size-prior-test}
\end{figure}

In order to study a wide range of possible galaxy kinematic configurations, we vary the PA in increments from 0 (perpendicular to the dispersion direction) to 90 (parallel to the dispersion direction) as shown schematically on Fig. \ref{fig:pa-cartoon}, then for each position angle we run the code for a grid of $V_a$ and $\sigma_0$ values: $V_a = 5, 100, 200, 300$ km/s and $\sigma_0 = 5, 25, 70, 100, 200, 500$ km/s. For the tests in this section, we set  $\text{SN}_{\text{max,grism}} = 20$.

Fig. \ref{fig:real_test_param} summarizes the results of these tests. When looking at $\sigma_0$ (top panel of Fig. \ref{fig:real_test_param}), we plot both the input and recovered values in logarithmic space in order to best show the evolution of the uncertainties in terms of dex. For the recovery of $\disp$, we see the same limitations at small $\disp$ as those highlighted in the idealistic tests (Sec. \ref{sec:ideal-test}), with boosted uncertainties due to the added noise and PSF/LSF convolutions in the mock grism observations. We note that even though the median of the posterior over-estimates $\sigma_0$ when it is well below the resolution, the width and shape of the posterior include the true value. In fact, in the $\sigma_0 = 5$ km/s, even though most of the uncertainties do not reach the true value, the posterior itself hits the $\sigma_0 = 0$ km/s lower bound. In \citet{Danhaive:2025aa}, we hence qualify these measurements as upper limits. The recovery of $v_{\rm re}$ (bottom panel of Fig. \ref{fig:real_test_param}) is good across the grid of parameters tested, although the errors are large when the velocity dispersion is at $\sigma_0 = 500$ km/s. However, this is not a normal value for normal star-forming galaxies, and is for testing purposes only.

To better visualize the accuracy of the recovery based on the input values and the PA, we plot on Fig. \ref{fig:real_test_param_mean_curves} the mean inferred values of $\disp$ and $v_{\rm re}$ for each PA. These mean values are computed by averaging over each $v_{\rm re}$ value for $\disp$, and vice-versa. Similarly to the idealised tests, we see that the $\rm PA = 90^{\circ}$ case does not allow for accurate recovery of the kinematic parameters due to the full degeneracy between the rotational velocities and the velocities dispersions. Otherwise, the recovery is good across the parameter space tested.

\subsubsection{Size test}\label{sec:morph-test}

We now focus on the morphology of the mock galaxies and how it affects our recovery of kinematic parameters. We note that the spatial pixel scale of the images used in these tests is $1 \rm \thinspace px = 0.063 ''$. For this test, we generate a mock observation with  $\text{SN}_{\text{max,grism}} = 10$, $\rm PA = 45^{\circ}$, $\disp = 100 \rm ~km/s$, and $v_{\rm re} = 170 ~ \rm km/s$. We then run \geko\ for a grid of \sersic\ indices $n=0.5,1,2,4,6,8$ and effective radii $r_{\rm e}\approx 2, 3,4,5,6$ pixels (derived from setting $r_{\rm t}\approx 0.25, 0.5,1,1.25,1.5$ pixels in Eq. \ref{eq:re-rt}).  The results are shown on Fig. \ref{fig:size-test}. We note that in this plot, the input kinematic values are fixed, but we spread them out artificially for clarity. 

Overall, we find that in most cases we are able to recover $\disp$ and $v_{\rm re}$ within the uncertainties. The most difficult scenarios are when galaxies have small sizes and specifically small turn-around radii ($r_{\rm t} \approx 0.15-0.25$), since this makes it very difficult to measure kinematics. This is worsened if the \sersic\ indices are large $n\geq 4$, since the light distribution is more centrally concentrated. In fact, the very large error visible on the top panel of Fig. \ref{fig:size-test}, and also present on the bottom panel, originates from the $n=6$ and $r_{\rm t} = 0.15$ extreme case. However, it is important to note that this test was run for a relatively low S/N case (S/N = 10) and the results naturally improve with higher S/N. Importantly, in nearly all of of the cases, the true values of $\disp$ and $\vre$ are included in the uncertainties.

The tests above are all done with morphological priors centred on their true value. We now explore how moving the priors away from the truth affects the recovery of both kinematic and morphological parameters. As shown on Tab. \ref{tab:tests}, we set $n_{\rm truth}=1$ and $r_{\rm e, truth}=4.19$ pixels for these tests, and instead vary the mean of the Gaussian prior for these parameters to test how much this biases the inferred posterior. We vary the \sersic\ index prior $\Delta n_{\rm prior} = n_{\rm prior} - n_{\rm truth} = -0.5, 1, 2, 3$, keeping the size prior fixed $\Delta r_{\rm e,prior} = 0$, then vary the size prior $\Delta r_{\rm e,prior} = -2,-1,1,2$ keeping the \sersic\ index prior fixed $\Delta n_{\rm prior} = 0$. For the recovery of $\disp$ and  $v_{\rm re}$, we show our results on Fig. \ref{fig:size-prior-test}. We find that we are able to recover the true values of $\disp$ and $v_{\rm re}$ within the uncertainties, but we note that the best-fit value of $v_{\rm re}$ gets worse when $\reff$ is underestimated (i.e. $\Delta r_{\rm e}<0$). This is most likely due to the fact that for small effective radii, it is more difficult to constrain the rotation curve out large enough radii. This was also shown on Fig. \ref{fig:size-test}, where for the smallest values of $\reff$, we were not able to recover $v_{\rm re}$ for $n\geq 4$.

\begin{figure}
    \centering
    \includegraphics[width=1\linewidth]{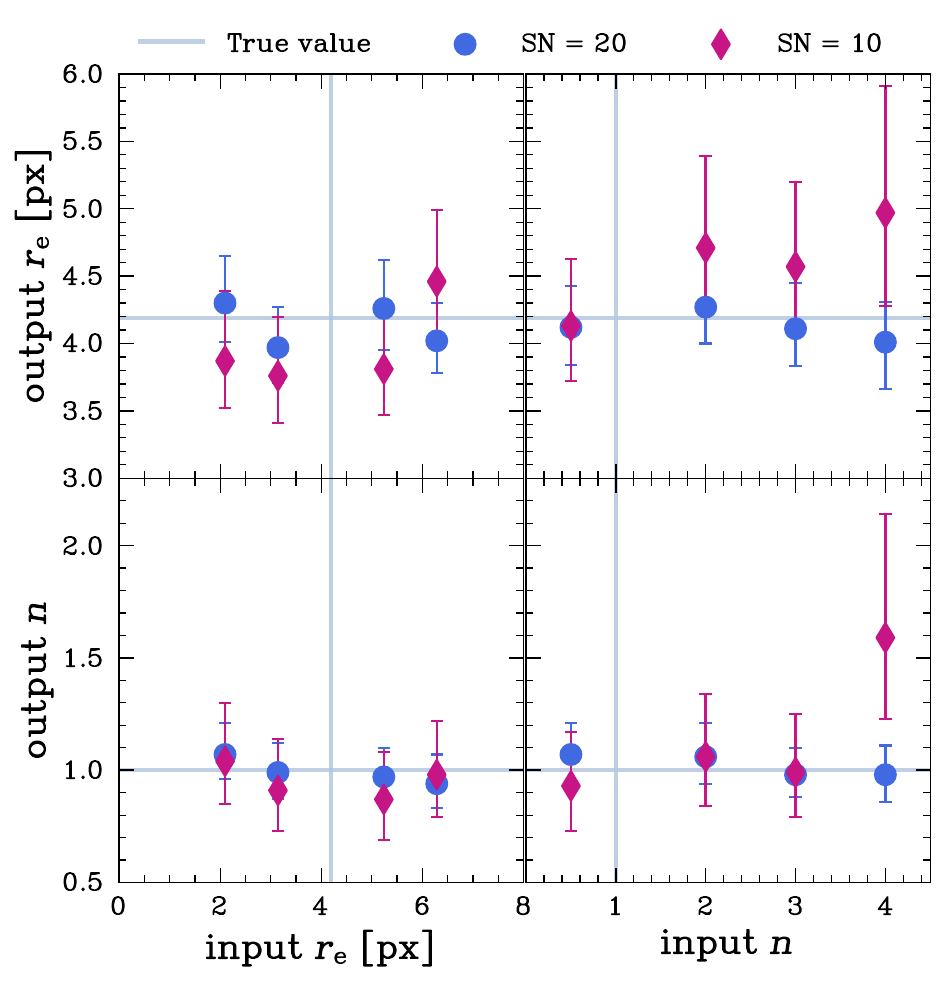}        
    \caption{Test results for the recovery of morphological parameters $\reff$ and $n$, using noisy mock grism data, when the mean of their priors is offset from the truth by $\Delta r_{\rm e}\approx -2,-1,1,2$ px and $\Delta n = -0.5, 1,2,3$. The true values are set to $r_{\rm e} \approx 4$ px and $n=1$ (blue solid line). When one parameter is varying, the other is fixed to their true value. The position angle of the mock galaxies is set to $\rm PA = 45^{\circ}$, and the kinematics to $V_a =200$ km/s and $\sigma_0 = 100$ km/s. The test is done for mock grism data with with $\rm S/N = 10$ (pink diamonds) and $\rm S/N = 20$ (blue circles).}
    \label{fig:prior-size-test}
\end{figure}

Even though the kinematic parameters are typically well recovered, recovering the right morphological parameters when the priors are offset from the truth is more difficult. This is in part due to a form of existing degeneracy between the \sersic\ index $n$ and the effective radius $\reff$. On Fig. \ref{fig:prior-size-test}, we show the recovery tests for these two parameters with $r_{\rm e,truth} \approx 4$ and $n_{\rm truth} = 1$, for mock grism data with $\rm S/N = 20$ and $\rm S/N = 10$. In terms of the recovery of $r_{\rm e,truth}$ (top panel of Fig. \ref{fig:prior-size-test}), we see that we are able to recover the true value well when $n_{\rm prior} = n_{\rm truth} =1$, particularly when $r_{\rm e,prior} < r_{\rm e,truth}$. The recovery is worse for the lower S/N test. This also holds for $n_{\rm prior} \neq n_{\rm truth}$, with the best-fit value getting worse with increasing $n$ and decreasing S/N.

For the recovery of $n$, (bottom panel of Fig. \ref{fig:prior-size-test}), we see that $n$ is almost always over-estimated, but remains consistent within the uncertainties. For $r_{\rm e, prior} = r_{\rm e, truth} \approx 4$, we are able to recover $n$ well even when the prior differs from the truth. The recovery is worse for the lower S/N case. For varying values of $r_{\rm e, prior}$, the best-fit value remains consistent but gets worse with decreasing $r_{\rm e, prior}$ and decreasing S/N.

This test case highlights the importance of the prior choice. We need to assume some of the emission line morphology, due to the morphology-kinematics degeneracy, and a wrong prior can bias results, despite having wide prior uncertainties. However, the recovery of kinematics (Fig. \ref{fig:size-prior-test}) remains robust.

\begin{figure}
    \centering
    \includegraphics[width=1\linewidth]{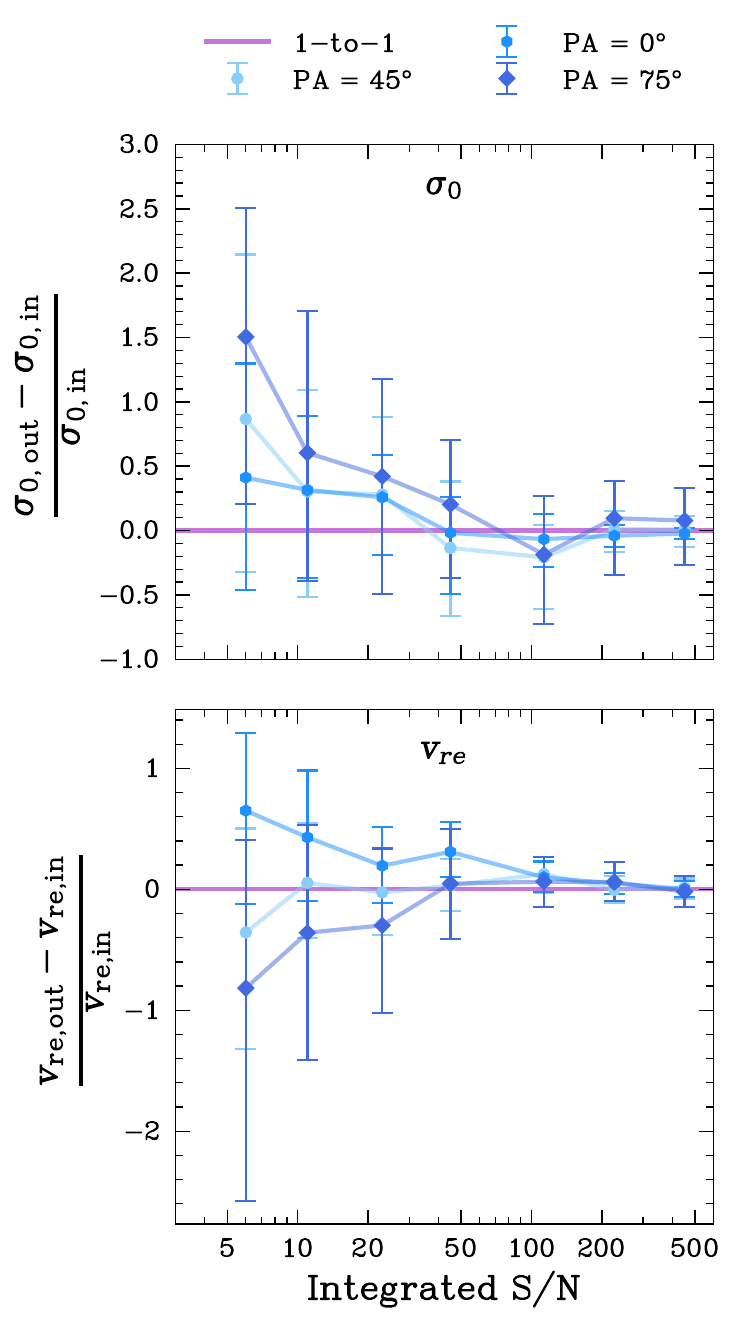}
    \caption{Evolution of the accuracy of the inferred $\disp$\ (top) and $\vre$ (bottom) values as a function of the integrated S/N of the mock grism data, for mock galaxies with position angles $\rm PA = 0^{\circ},45^{\circ},75^{\circ}$, and kinematics defined by $V_a =200$ km/s (i.e., $v_{\rm re} \approx 170$ km/s) and $\sigma_0 = 100$ km/s. }
    \label{fig:sn-test}
\end{figure}

 \begin{figure*}
    \centering
\includegraphics[width=1\linewidth]{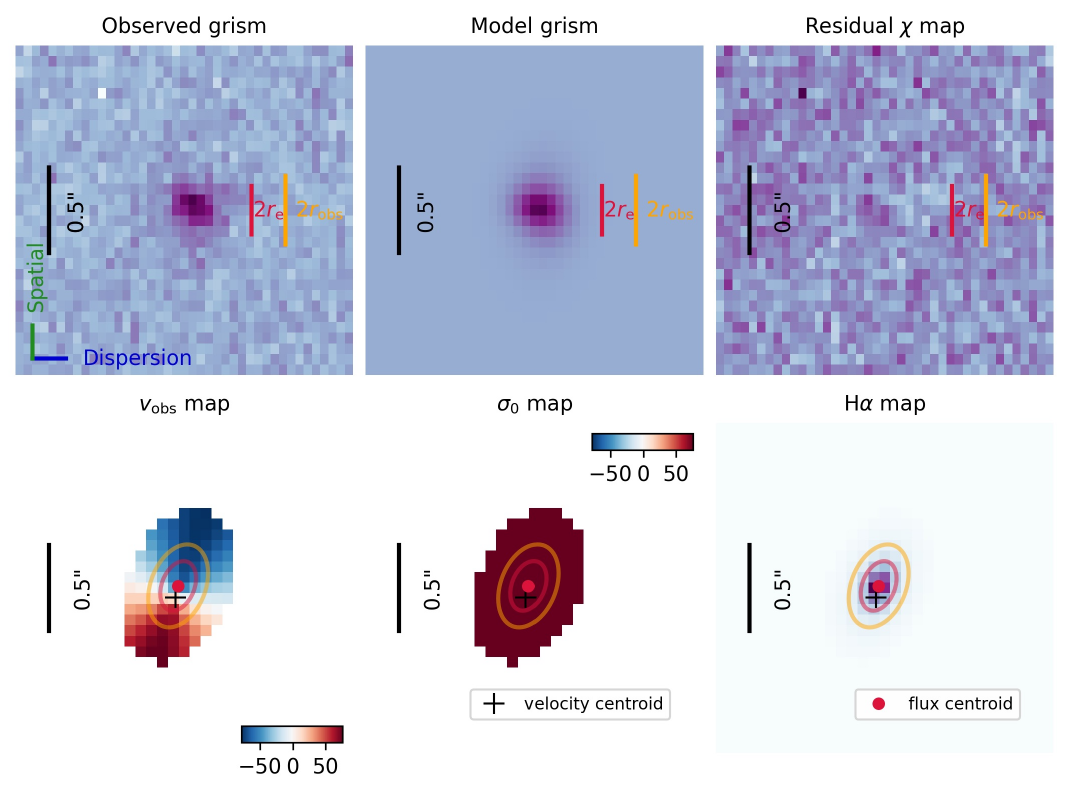}
    \caption{Example science products obtained with \geko\ for a galaxy in the FRESCO survey (JADES ID 1025527). In the top panels, we show the observed grism data, the best-fit model, and the corresponding residuals computed with the $\chi$ metric. The plotting range is $\chi = [-5, - 5]$. We also highlight the best fit effective diameter $D_{\rm e} = 2r_{\rm e}$ for the \Ha\ emission, and the (projected) diameter of the $\text{S/N}>3$ observed map. In the bottom panel, we show the derived best-fit velocity field, velocity dispersion field (which has the same velocity scale as the velocity field plot), and the intrinsic \Ha\ emission map. In all of the panels on the left side, we show the scale in arcseconds (black bar) along with the velocity centroid (black cross). In each row, we show the physical meaning of the axes (spatial vs dispersion).}
    \label{fig:geko-summary}
\end{figure*}

\subsubsection{S/N test}\label{sec:sn-test}

We now present our recovery tests of kinematic parameters as a function of the integrated $\rm S/N$, for a mock galaxy with $\disp = 100 \rm ~ km/s$, $v_{\rm re} \approx 170 \rm ~ km/s$, and for position angles $\rm PA = 0^{\circ}, 45^{\circ}, 75^{\circ}$ (see Fig. \ref{fig:pa-cartoon}). We show our results on Fig. \ref{fig:sn-test}. Starting with $\disp$, we first see that, as expected, our best-fit values converge around the true values for high $\rm S/N$ values. This is consistent with our mock tests on idealized noiseless data (Sec. \ref{sec:ideal-test}). Also, the uncertainties decrease with increasing S/N, with the best-fit value approaching the true value. Below $\rm S/N = 10$, the recovery is not satisfactory for the worst "allowed" alignment, $\rm PA = 75^{\circ}$, given the LSF. However, above this S/N threshold, we recover the input $\disp$ within 1$\sigma$, despite over-estimations which are more pronounced $\rm PA = 75^{\circ}$ . 

For $v_{\rm re}$, we find similar results, and note that for the $\rm PA = 0^{\circ}$ case, where the galaxy is perpendicular to the dispersion direction, \geko\ recovers the true value well even at low S/N, with an uncertainty that decreases with increasing S/N. This test further highlights the benefit of having galaxies with an orientation as distant as possible from the dispersion direction.

\section{Application  to real data} \label{sec:real-data}

We now demonstrate the usage of \geko\ on real data, using as examples various \Ha\ emitters at $z\approx 4-6$ taken from the sample described in \citet{Danhaive:2025aa}. In Sec. \ref{sec:data}, we present the data used in the fitting and, in Sec. \ref{sec:morph-fit}, we describe the morphological modelling with \textsc{pysersic} to construct a prior for the kinematic modelling. In Sec. \ref{sec:results}, we show \geko\ fitting results for some representative galaxies. 

\subsection{Data}\label{sec:data}
The grism data for our objects are taken from the \textit{JWST}/NIRCam WFSS observations in the F444W filter ($3.9-5.0 \thinspace\mu$m) in GOODS-N obtained with the FRESCO \citep{Oesch:2023aa,Covelo-Paz:2025aa} survey. The observations are taken using the row-direction grisms (GRISMR mode) with modules A and B, and the spectral resolution is $R\approx 1600$. The data reduction for these observations is described in detail in \cite{Sun:2023aa} and \cite{Helton:2024ab}. 

The FRESCO footprint has a large overlap with JADES \citep{Eisenstein:2023aa,Rieke:2023aa}, a Guaranteed Time Observations (GTO) programme of the NIRCam and NIRSpec instrument teams that obtained imaging of an area of $\sim 175$ arcmin$^2$ in the GOODS-S and GOODS-N fields with an average exposure time of 20 hours with 8-10 filters. For this galaxy, we were able to complement the FRESCO F182M and F210M imaging with wide band F090W, F115W, F150W, F200W, F277W, F356W, and deeper F444W observations, and medium band F335M and F410M observations from JADES. Specifically, we use high-resolution data and photometric catalogues obtained from drizzled mosaics, and the full details on the data reduction, catalogue generation, and photometry can be found in \citet{Rieke:2023aa}, and are also detailed in \citet{Robertson:2024aa}. The wealth of photometric data is used for SED modelling of this system, which is not discussed here \citep[see][]{Danhaive:2025aa}, but also to model its morphology and infer physically-motivated prior for the inference process.

\subsection{Morphological fitting}\label{sec:morph-fit}

An important part of using \geko\ with a parametric flux model (Sec. \ref{sec:morph}) is to is to chose an informed prior (Sec. \ref{sec:priors}). In order to obtain constraints on the morphological parameters of our galaxies, we use the fully Bayesian code \textsc{pysersic} \citep{Pasha:2023aa}. \textsc{pysersic} models imaging data from any filter to infer the best fit \sersic\ profile(s) parameters. 

We run \textsc{pysersic} on the F150W imaging of our system, because at $z\approx 5.3$, F150W probes the near-UV continuum ($\lambda_{\rm rest} \sim 2000 \rm ~\mathring{A}$). We fit the image of our galaxy with a one-component \sersic\ model for every galaxy in our sample to obtain estimates of the position angle $\theta$, ellipticity $e$, \sersic\ index $n$, light centroid $(x_0,y_0)$, and half-light radius $\reff$. We note that prior to fitting, the images from JADES (Sec. \ref{sec:data}) are PA-matched to the grism data since the imaging and grism come from different surveys. 
Also, as described in Sec. \ref{sec:morph}, we assume an intrinsic axis ratio $q_0=0.2$ when deriving the inclination from the ellipticity. Because of how we define the position angle from the vertical y-axis, we note the conversion $\text{PA}_{\text{morph}} = 90 - \theta$ degrees, where $\theta$ is the \textsc{pysersic} output angle computed from the positive x-axis.

We estimate of the total \Ha\ flux from the integrated 1D grism spectrum. We use Gaussian priors for all of these morphological parameters, doubling the uncertainty obtained from \textsc{pysersic} since we are not measuring the true emission-line morphology, but instead stellar continuum from star-forming regions. For the kinematic parameters, we use uniform priors on $\disp$, $V_a$, and $r_t$, where this last parameter is bounded by \reff, motivated by lower redshift studies finding $r_t \sim 0.25 \thinspace r_{\rm e}$ \citep{Miller:2011aa}. The prior for the kinematic position angle \PAkin is the same as for \PAmorph. The free parameters of our model and their priors are summarized in Tab. \ref{tab:priors}.

\subsection{Results}\label{sec:results}

\begin{figure}
    \centering
\includegraphics[width=1\linewidth]{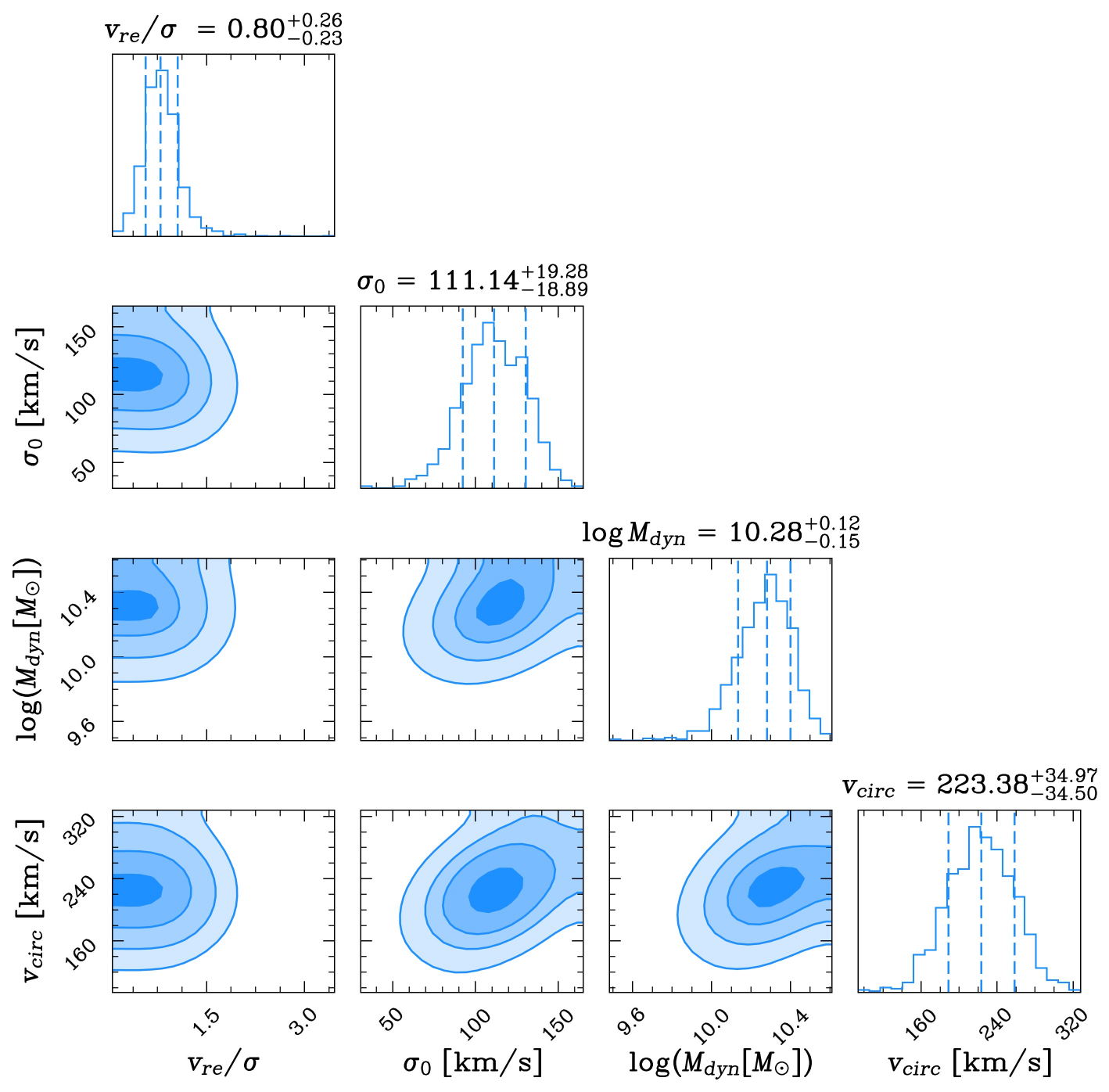}
    \caption{Posterior distributions inferred with \geko\ for the galaxy from Fig. \ref{fig:geko-summary}, for the key model-derived parameters. The uncertainties are computed from the $16^{\rm th}$ and $84^{\rm th}$ quantiles, are shown on the too of the marginalised posteriors.}
    \label{fig:vsigma-corner}
\end{figure}

\begin{figure*}
    \centering
    \includegraphics[width=1\linewidth]{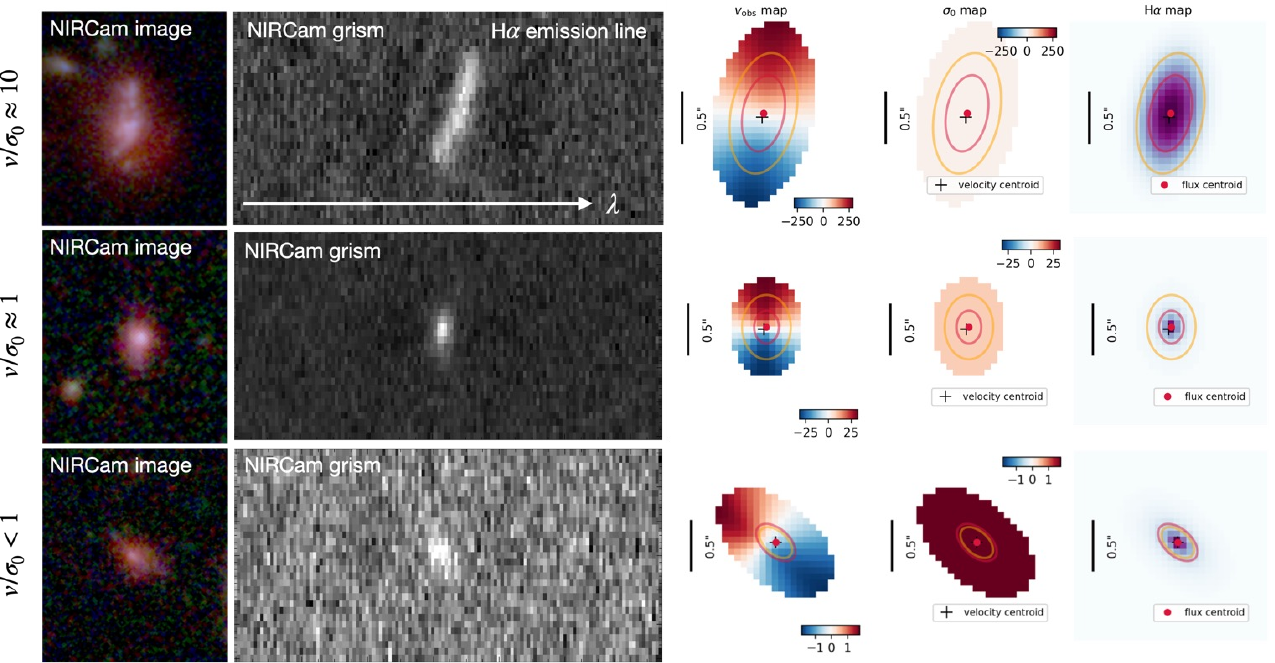}
    \caption{(From left to right) RGB image cutout from the JADES mosaics, NIRCam grism data from FRESCO, and \geko\ fit results for three galaxies at $z\approx 5$ with significant ($\rotsupp\approx 10$), marginal ($\rotsupp\approx 1$), and no ($\rotsupp<1$) rotational support. }
    \label{fig:galaxies-examples}
\end{figure*}

Using the priors set from the morphological fitting (Sec. \ref{sec:morph-fit}), we run \geko\ to constrain the emission line morphology and kinematics for our chosen example galaxies. We show the full \geko\ fit summary on Fig. \ref{fig:geko-summary} for an \Ha\ emitter at $z=5.295$. This summary figure is directly output from the \geko\ post-processing pipeline (Sec. \ref{sec:postpro}) in order to facilitate the visualisation of the fit by the user. On the top left plot of Fig. \ref{fig:geko-summary}, we show the observed continuum subtracted grism data centred on \Ha, followed by our best fit model and the residual $\chi$ map which is defined by:
\begin{equation}
    \chi = \frac{\text{model} - \text{obs}}{\text{obs uncertainty}},
\end{equation}
and where the bounds for plotting are $\chi = [-5, 5]$. On the bottom row we show the inferred velocity, intrinsic velocity dispersion, and intrinsic flux maps for \Ha. We note these maps map are PSF-deconvolved. The plotted observed velocity $v_{\rm obs}$ is related to the source-plane rotational velocity through Eq. \ref{eq:vobs}. The centre of the velocity curve is marked by a black cross on all three plots, whereas the flux centroid is marked by a red dot. These centroids are fit independently and are free to be distinct, as shown by the small offset between the two. We also show the comparison between the half-light radius, as inferred from the \geko\ modelling, and the observed extent of the emission $r_{\rm obs}$. Typically, it is best to require that the observed emission line probes the emission at least out to \reff ($r_{\rm obs}>$\reff), so that the measured rotational velocity at the effective radius ($\vre$) is well constrained. On Fig. \ref{fig:geko-summary}, we can see that this criteria is verified in this fit. 

On Fig. \ref{fig:vsigma-corner}, we show the posterior distributions for the key inferred kinematic parameters: the rotational support $\rotsupp$, the velocity dispersion $\disp$, the dynamical mass \Mdyn, and the circular velocity \vcirc. This plot also constitutes an output of the \geko\ post-processing pipeline. Except for the velocity dispersion, the rest of these quantities are not direct parameters of the model, but instead are derived in the post-processing following the definitions from Sec. \ref{sec:postpro}. This object shows some rotational velocity, but is nonetheless dominated by its elevated velocity dispersion $\disp = 111 \pm 20$ km/s, yielding $\rotsupp = 0.80\pm 0.25$. This is not surprising at high redshift, where galaxy formation is expected to be turbulent.

On Fig. \ref{fig:corner} we show the posterior distributions of all of the free parameters of the model, which is also an output of the \geko\ post-processing. We also highlight the best fit values and uncertainties, computed from the $16^{\rm th}$ and $84^{\rm th}$ quantiles, for the three key kinematic parameters studied in this work: the intrinsic velocity dispersion $\disp$, the rotational velocity at the effective radius $v_{\rm re}$, and the rotational support $\rotsupp$. Fig. \ref{fig:corner} highlights the tight kinematic posteriors despite the flat priors, indicative of the information encoded in the grism data. The size $\reff$ and the \sersic\ index are also well constrained. The inference model does not learn any additional information regarding the morphological position angle, as can be seen by the posterior tracing the prior for $\rm PA_{morph}$. This is due to the tightness of the prior, which is necessary to not bias the kinematic measurements. However, the kinematic PA is left more free to vary. The model is also not able to offer tighter constraints on the centre of rotation in this case.

We now showcase some more example of fitting galaxies with \geko. On Fig. \ref{fig:galaxies-examples}, we show examples of three \geko\ fits for three systems with varying degrees of rotational support. For each system, we show the NIRCam RGB image, obtained from mosaics from the JADES survey, the grism 2D emission line map, and the best-fit intrinsic maps from \geko. The mosaics have a 0.031'' pixelisation, consistent with the NIRCam short-wavelength resolution, whereas the grism data is taken in the long-wavelength filters, and hence has a 0.063'' pixelisation. In all three cases, the grism spectra are centred on the \Ha\ line. On the top panel, we show an \Ha\ emitter at $z=5.2$, which was first reported as a highly rotating system in \citet{Nelson:2024aa}. Our modelling with \geko\ yields a high rotational velocity $v_{\rm rot}(r_{\rm e})= 231^{+21}_{-15}$ km/s, and low velocity dispersion $\sigma_0 = 21 \pm 18$.  In \citet{Nelson:2024aa}, they report a rotational velocity of $v_{\rm rot}(r_{\rm e})= 240^{+50}_{-50}$ km/s. This object is also discussed in \citet{Li:2023aa} where they find a similar value of $v_{\rm rot}(r_{\rm e})= 211^{+21}_{-18}$ km/s. Our best fit value is in good agreement with these works. However, both \citet{Nelson:2024aa} and \citet{Li:2023aa} find relatively low values for the rotational support, $\rotsupp\sim2$, whereas we obtain a higher value of $\rotsupp\sim 10$. This is likely due to differences in the morphological and kinematic modelling. For example, \citet{Nelson:2024aa} do not forward model the impact of the LSF but subtract it in quadrature, and \citet{Li:2023aa} match the image PSF to the LSF instead of modelling their effects separately. 

In the middle panel, we show a galaxy with equal contributions from the rotational support and the velocity dispersion ($\rotsupp\approx 1$). In fact, we measure relatively low velocities ($v_{\rm rot}(r_{\rm e})= 41^{+15}_{-13}$ km/s), but also low values of velocity dispersion ($\sigma_0 = 27 \pm 20$ km/s), resulting in $\rotsupp = 1.5^{+3.7}_{-0.8}$. Finally, in the bottom panel, we present a galaxy that is dispersion dominated ($\rotsupp< 1$). This galaxy has no resolved velocity gradient, and a high intrinsic velocity dispersion ($\sigma_0 = 82.6^{+48}_{-51}$ km/s). This implies that the ionised gas in this system, probed by the \Ha\ emission, is subject to high degrees of turbulence. 

In this section, we have shown applications of \geko\ to real galaxies from the FRESCO survey (Fig.~\ref{fig:galaxies-examples}). These galaxies are taken from the large sample analysed, which is analysed in \citet{Danhaive:2025aa}. In that work, we measure the ionised gas kinematics of $\approx 200$ \Ha\ emitters from the FRESCO survey and the the COmplete Nircam Grism REdShift Survey (CONGRESS, PIs: Egami, Sun, PID: 3577) survey, which is limited to GOODS-N. Using SED modelling with \textsc{Prospector} \citep{Johnson:2021aa}, we constrain the stellar masses and star-forming properties of the galaxies in our sample, and relate them to the inferred kinematics.

\section{Prospects for future development} \label{sec:discussion}

The present release of \geko\ demonstrates the feasibility of recovering emission-line morphologies and kinematics from \textit{JWST}/NIRCam slitless spectroscopy using forward modelling and Bayesian inference. Several avenues for further development are planned. First, as mentioned in Sec. \ref{sec:morph}, we are in the process of implementing the morphological fitting on the imaging data in the code itself. We currently fit using \textsc{Pysersic} prior to running \geko. but also allow for morphology inputs from the user.

Second, within our simple kinematic and morphological models, we plan to implement multiple component fitting. A two-component approach would be important for fitting mergers, as well as clumpy systems, especially if the clumps have differing kinematics. Also, including a model for characterizing active galactic nuclei (AGN) would broaden the use of \geko\ and help study the kinematics of AGN at high redshift. In particular, such modelling would require the addition of a broad line region (BLR). This two-component approach would allow us to fit the broad and narrow emission of Type-I AGN.

Furthermore, \geko\ currently only fits one emission line at a time. However, many grism surveys are obtaining grism data in different filters, probing various emission lines for single objects. Although one can currently model each line and then combine the posterior distributions, we plan to implement the joint fitting of emission lines in \geko\ to improve the derived kinematics constraints and, importantly, their uncertainties. 

Finally, and most importantly, with the arrival of grism data taken with both dispersion directions \citep[e.g. SAPPHIRES;][]{Sun:2025aa}, namely R and C, we will implement the joint fitting of these data in order to recover intrinsic maps and velocity gradients with more flexibility. In fact, because the R and C grism modes disperse the light in perpendicular directions, we are able to further alleviate the morphology-kinematics degeneracy. Assuming a simple kinematic model, we will be able to reconstruct a spatially resolved emission line map without having to assume a \sersic\ light distribution. This would, for example, allow for studies of resolved gradients when the necessary emission lines are available. In the same way, this methodology applies to any grism data obtained with multiple different position angles. Even a small difference can allow for additional constraints in the emission line morphology.

\section{Conclusions}
With the arrival of the \textit{JWST}, and specifically the grism mode of the NIRCam instrument, we are able to study large samples of galaxies with high spectral and spatial resolution. These new data present exciting opportunities, but also require novel analysis tools to extract the relevant information. Given the large sample sizes, such tools must not only be robust but also be computationally efficient.

We have developed a grism kinematics and morphology modelling tool called \geko, whose aim is to constrain the rotational velocities, velocity dispersions, and sizes of galaxies by probing emission lines across redshifts. By leveraging the benefits of \textsc{numpyro} and \textsc{jax} programming, \geko\ is optimised for GPU usage, where it is able to fit a galaxy in $\sim 10-30$ min. In this paper, we present a first version of \geko, and release the code for public use. We summarize the key elements of \geko\ here. 

\geko\ fits the emission line morphology and kinematics using parametric models, namely a \sersic\ profile and an arctangent velocity curve, which are projected onto the sky based on the inclination and position angle of the galaxy. Each parameter in the two models is left free to vary, with the morphological ones having Gaussian priors to help break the degeneracy with the kinematic imprint on the 2D grism spectrum. Beyond obtaining posteriors for the model parameters, \geko\ includes post-processing which infers key derived properties such as the rotational support and dynamical mass. It also produces visualizations of the posteriors and best fit models and residuals to facilitate scientific interpretation. As demonstrated in \citet{Danhaive:2025aa}, this tool opens the door to statistical analyses of galaxy kinematics at high redshift, ushering in the study of early galaxy formation through a dynamical lens.


\section*{Acknowledgements}

We thank Benjamin D. Johnson, Fengwu Sun and Eiichi Egami for technical insights which helped in the development of the code. We thank Hannah Übler, Emily Wisnioski, William McClymont and the JADES collaboration for helpful discussions. We thank Amanda Stoffers for the design of the GEKO logo. ALD thanks the University of Cambridge Harding Distinguished Postgraduate Scholars Programme and Technology Facilities Council (STFC) Center for Doctoral Training (CDT) in Data intensive science at the University of Cambridge (STFC grant number 2742605) for a PhD studentship. ALD and ST acknowledge support by the Royal Society Research Grant G125142.






\bibliographystyle{mnras}
\bibliography{main} 



\newpage
\appendix

\section{Coordinate transformations}\label{sec:annexe_vel}
\begin{figure*}
    \centering
    \includegraphics[scale = 0.3]{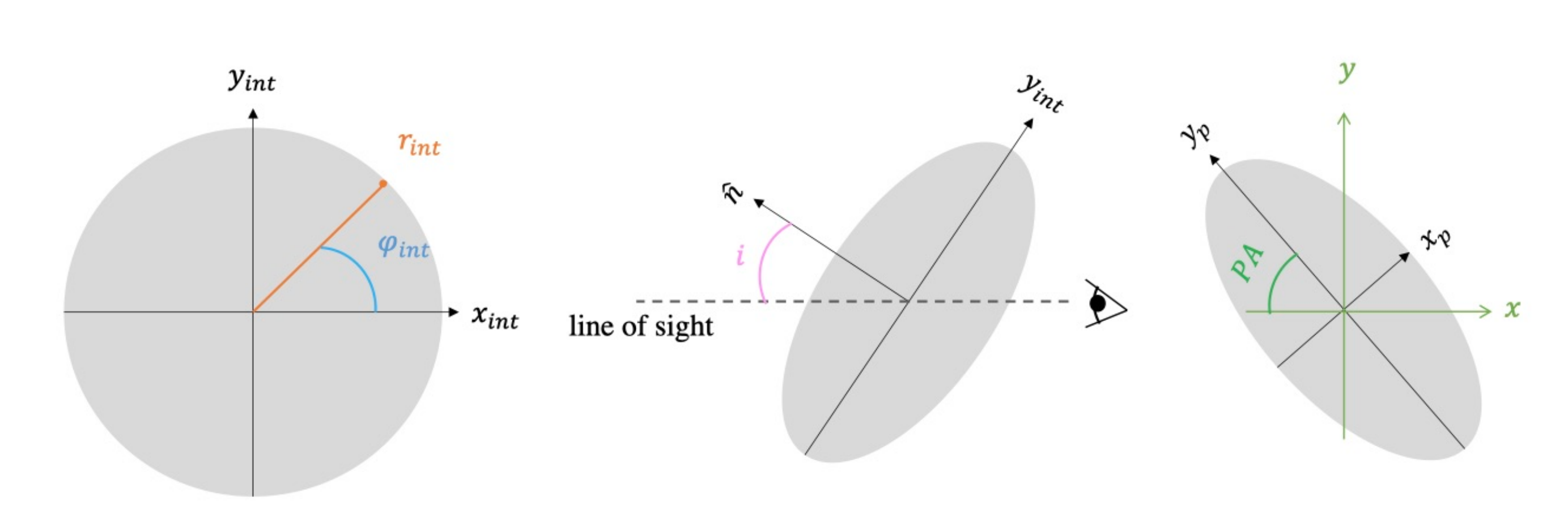}
    \caption{Sketch of the variables involved in the coordinate transformation from intrinsic to observed galaxy coordinates (Eq. \eqref{eq: coord_trans} ). Here, $x_p,y_p$ are the projected coordinates obtained by inclining the galaxy: $x_p = x_{\text{int}},y_p =y_{\text{int}}\cdot \cos{i} $. }
    \label{fig:coordinate_schema}
\end{figure*}

In the section, we detail the derivation of the observed galaxy coordinates defined in Eq. \ref{eq: coord_trans} from the galaxy plane. This also gives insight into how the angles are defined within \geko. On Fig. \ref{fig:coordinate_schema}, we show this process through a visual diagram. On the left, we show the galaxy plane, where $r_{\text{int}}$ is the distance from the galaxy center, and $\phi_{\text{int}}$ is the angle from the x-axis.
The intrinsic coordinates on the galaxy plane are therefore:
\begin{align}
    x_{\text{int}} =& r_{\text{int}}\cos\phi_{\text{int}} \\
    y_{\text{int}} =& r_{\text{int}}\sin\phi_{\text{int}}.
\end{align}
To project the galaxy on the sky, we first need to incline it. We define the inclination as the angle from the line of sight, such that $i=0^{\circ}$ corresponds to a face on galaxy, and $i=0^{\circ}$ to an edge-on galaxy. On the sky, the projected coordinates become:
\begin{align}
    x_p =& x_{\text{int}}  \\
    y_p =& y_{\text{int}} \cdot\cos i.
\end{align}
Finally, we need to rotate the galaxy to reflect its position angle, meaning the angle between the galaxy's major axis and the x-axis, as defined on the sky. This effectively rotates the coordinates:
\begin{align}
    x_{\text{obs}} =& x_{\rm p}\cos{\rm -PA} - y_{\rm p}\sin{\rm -PA} = x_{\rm p}\cos \rm PA + y_p\sin \rm PA   \\
    y_{\text{obs}}=& x_{\rm p}\sin{\rm -PA} + y_{\rm p}\cos{\rm -PA} = -x_{\rm {\rm p}}\sin \rm PA + y_{\rm p}\cos{\rm PA}.
\end{align}
We have now derived the relation between the coordinates on the sky and those on the intrinsic galaxy plane:
\begin{align}
    x_{\text{int}} = x_{\rm p} =  x_{\text{obs}}\cos{\rm PA} - y_{\text{obs}}\sin{\rm PA}\\
    y_{\text{int}}\cos{i} = y_{\rm p} = x_{\text{obs}}\sin{\rm PA} + y_{\text{obs}}\cos{\rm PA}.
    \label{eq: coord_trans}
\end{align}
This allows us to move between the two reference frames, and in particular convert from the rotational velocity $V_{\text{rot}}$, as defined on the galaxy plane, to the measured observed velocity $V_{\text{obs}}$:

\begin{equation}
    V_{\text{obs}}(x_{\text{obs}}, y_{\text{obs}}) = V_{\text{rot}}(r_{\text{int}})\cdot \cos{\phi} \cdot\sin{i}.
\end{equation}
These equations are implemented in \geko, and are optimised using just-in-time (jit) compilation since they form the basis of the core functions.

\section{Additional validation and test details}\label{app:ideal-tests}

This section presents additional details and figures that complement the testing described in the paper. 

\begin{figure}
    \centering
    \includegraphics[width=1\linewidth]{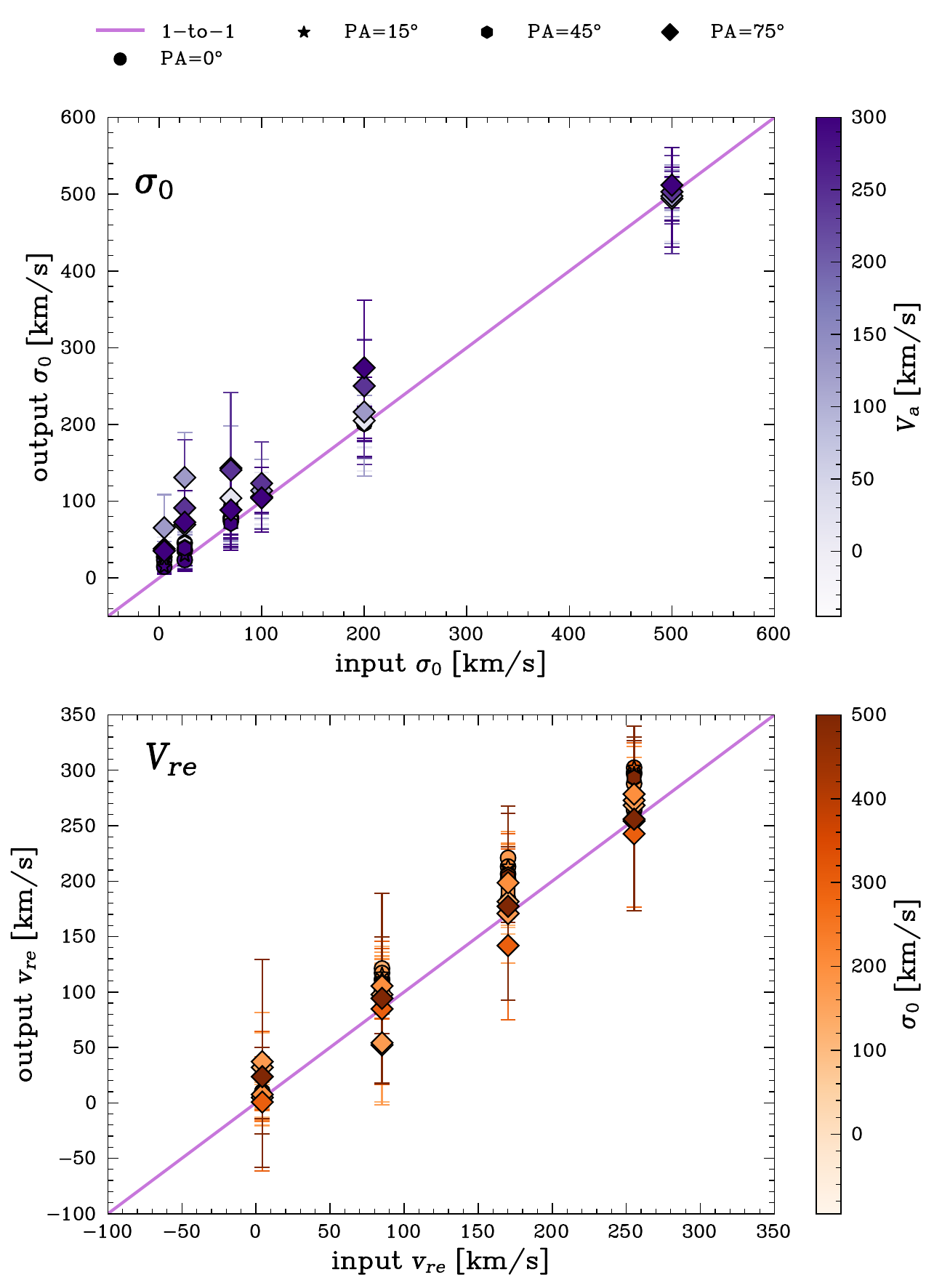}
    \caption{Test results for the recovery of kinematics, using noiseless mock grism data, for a range of position angles, $\rm PA = 0^{\circ},15^{\circ},45^{\circ},75^{\circ}$, asymptotic velocities $V_a = 5,100,200,300$ km/s, and velocity dispersions $\sigma_0 = 5,25,70,100,200,500$ km/s. For clarity, we do not show the $\rm PA = 90^{\circ}$ where $\sigma_0$ and $V_a$ are degenerate (see Fig. \ref{fig:PA90-deg}).}
    \label{fig:ideal_test_param}
\end{figure}

Figure~\ref{fig:ideal_test_param} shows the results of the idealized kinematic recovery tests performed on noiseless mock grism data, presented in Sec. \ref{sec:ideal-test}. These tests explore a broad parameter space in asymptotic velocity ($V_a$), intrinsic velocity dispersion ($\sigma_0$), and position angle (PA), quantify the recovery of input parameters for the idealised case. They demonstrate that the model reliably recovers the true kinematic parameters over a wide range of physical conditions, except for the case of $\rm PA = 90^{\circ}$, where the degeneracy between the rotational and dispersion components becomes dominant. We find that \geko\ cannot recover extremely low velocity dispersions $\disp$, as is expected from the relatively broad LSF of the grism. However, for higher values, the recovery is reliable for low PAs, and is worsened for PAs closer to the dispersion direction ($\rm PA = 90^{\circ}$). 

\begin{figure}
    \centering
    \includegraphics[width=1\linewidth]{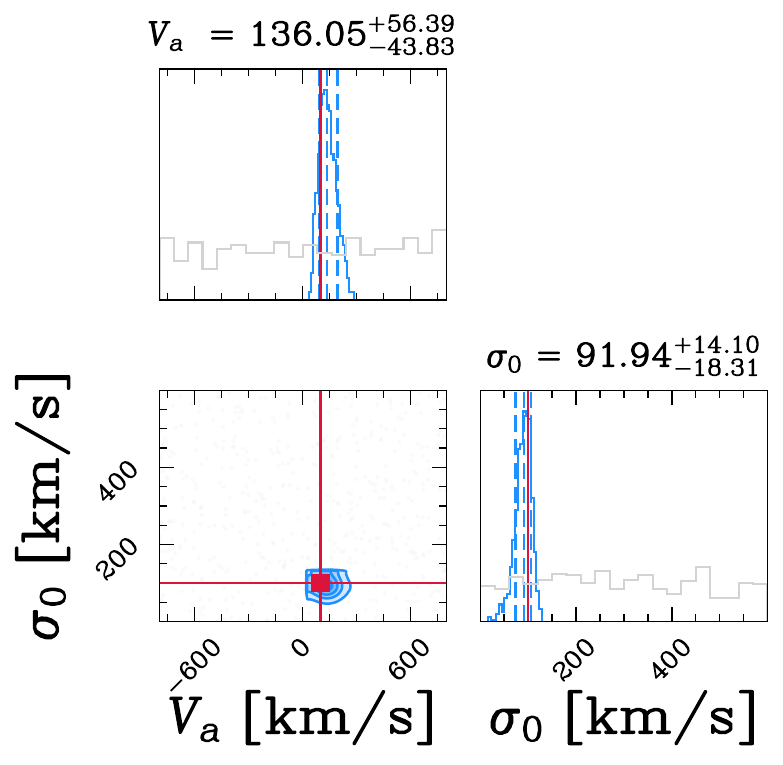}
    \caption{Posterior distributions for $\disp$\ and $V_a$ for a mock galaxy with $PA = 0$ degrees, $V_a = \sigma_0= 100$  km/s. The true parameters are recovered accurately and there is no visible degeneracy between the two parameters.}
    \label{fig:PA0-deg}
\end{figure}

To explore the bad recovery in the $\rm PA = 90^{\circ}$ case, we show the posterior distributions for $V_a$ and $\disp$ in the cases of $\rm PA = 0^{\circ}$ (the galaxy is perpendicular to the dispersion direction; Fig. \ref{fig:PA0-deg}) and $\rm PA = 90^{\circ}$ (the galaxy is parallel to the dispersion direction; Fig. \ref{fig:PA0-deg}). While the parameters are accurately recovered and uncorrelated in the $\rm PA = 0^{\circ}$ case, a clear degeneracy appears at $\rm PA = 90^{\circ}$, confirming the expected limitations when the galaxy is aligned with the dispersion direction. For this reason, in \citet{Danhaive:2025aa} we enforce at least $\rm PA < 75^{\circ}$ to mitigate this effect. 

Moving to the realistic tests described in Secs. \ref{sec:real-tests}, we present our testing strategy on Tab.~\ref{tab:tests}. Tab.~\ref{tab:tests} summarizes the different sets of tests carried out, detailing the parameters that were varied and those kept fixed in each experiment. These are also described in the relevant sections for each test.

\begin{figure*}
    \centering
\includegraphics[width=1\linewidth]{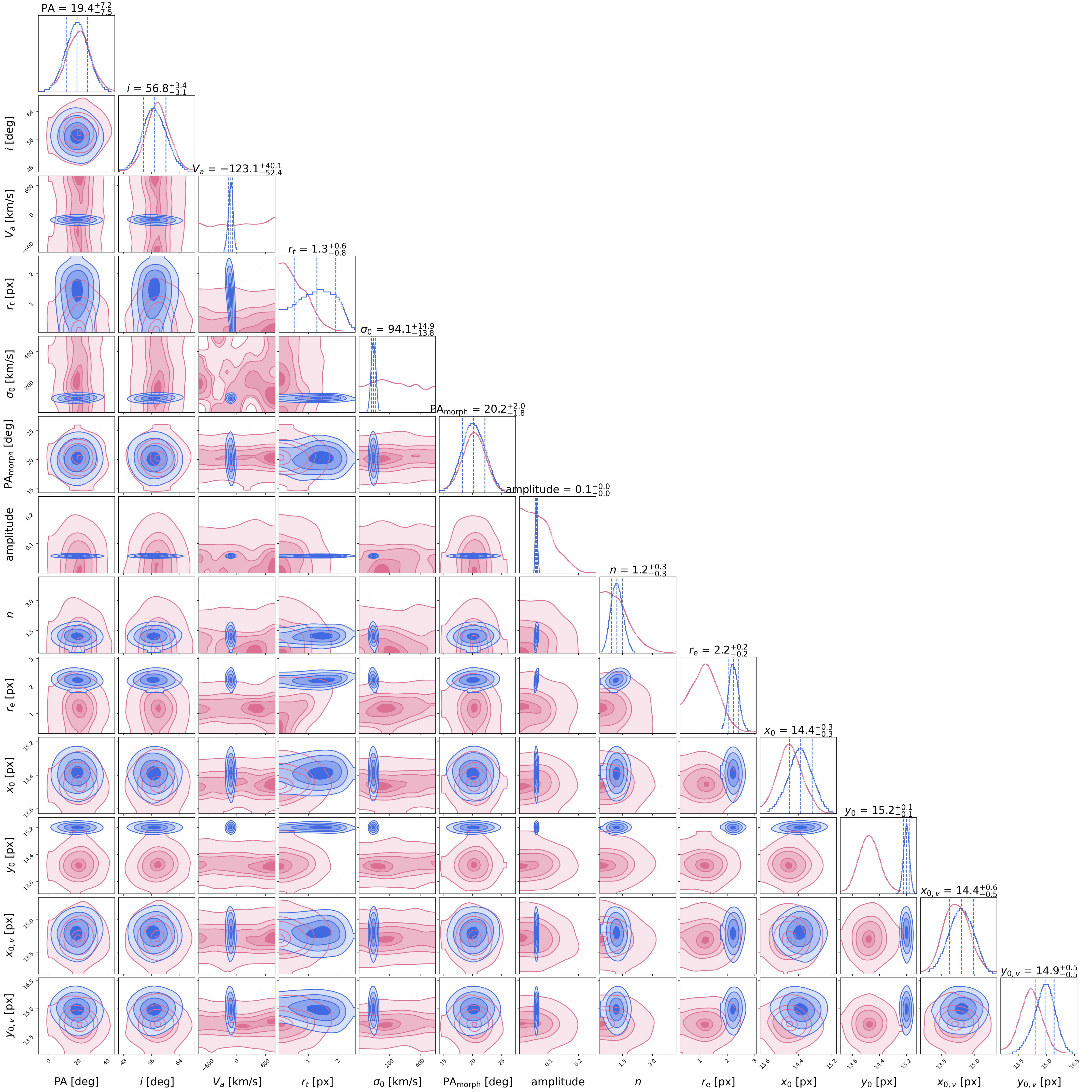}
    \caption{Prior (pink) and posterior (blue) distributions inferred with \geko\ for the JADES galaxy from Fig. \ref{fig:geko-summary}, for the free parameters of our model (Tab. \ref{tab:priors}). From these, we derive the posterior distributions for the rotational velocity at $v_{\rm re}$ and the rotational support $\rotsupp$, whose best fit values and uncertainties, computed from the $16^{\rm th}$ and $84^{\rm th}$ quantiles, are shown on the top right corner.}
    \label{fig:corner}
\end{figure*}

Finally, Fig.~\ref{fig:corner} presents the corner plot of the posterior distributions inferred with \geko\ for the example galaxy shown on Fig. \ref{fig:geko-summary}. The posteriors are shown in blue and the priors in pink. The priors for the kinematic parameters are uniform, and the posteriors are well constrained for $\disp$ and $V_a$. The turn-around is not as well constrained, which is expected. The morphological posteriors are informed from the priors, but show narrower posteriors for the total emission-line flux, \sersic\ index, and effective radius. On the other hand, the position angle and inclination have tighter priors and the posteriors are hence prior-dominated.

\begin{figure}
    \centering
    \includegraphics[width=1\linewidth]{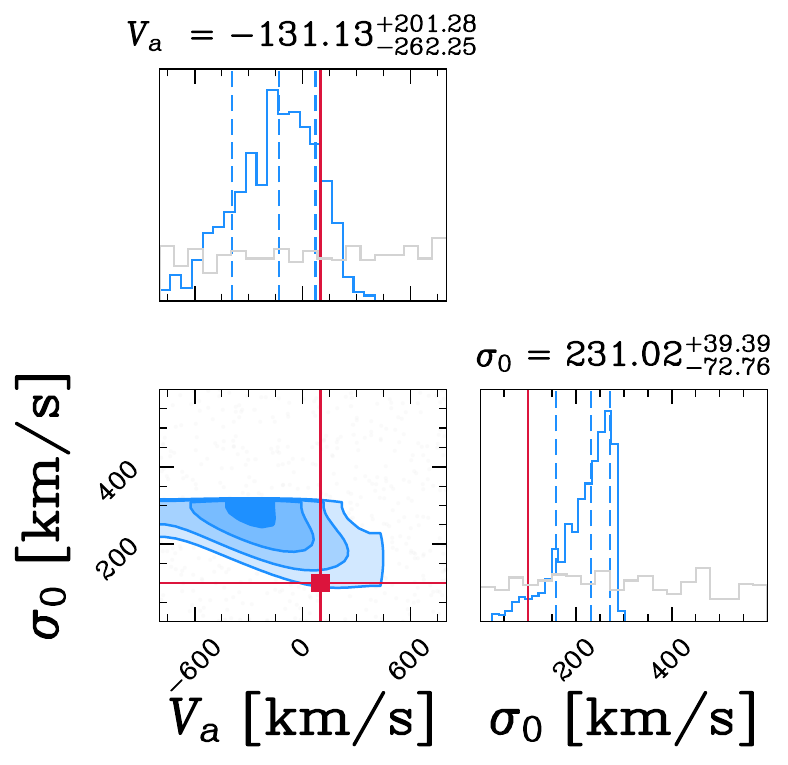}
    \caption{Posterior distributions for $\disp$\ and $V_a$ for a mock galaxy with $PA = 90$ degrees, $V_a = \sigma_0= 100$  km/s. Contrary to Fig. \ref{fig:PA0-deg}, here a strong degeneracy appears between the two parameters, caused by the fact that the galaxy is now parallel to the dispersion direction. This also causes large uncertainties in the estimates obtained.}
    \label{fig:PA90-deg}
\end{figure}

\begin{table*}
\centering
\begin{tabular}{c|c|c}
Name                             & Varying parameters                    & Fixed parameters                                                                                                                                                                                    \\ \hline
\multirow{3}{*}{Kinematics test (Sec. \ref{sec:kin-test})} & $\rm PA = 0, 15, 45, 75, 90^{\circ}$  & \multirow{3}{*}{\begin{tabular}[c]{@{}c@{}}$\rm S/N = 20$, $i = 60^{\circ}$,\\  $r_t = 1$, $r_{\rm e} = 4.19$, $n=1$\end{tabular}}                                                                  \\ \cline{2-2}
                                 & $V_a = 5, 100, 200, 300$ km/s         &                                                                                                                                                                                                     \\ \cline{2-2}
                                 & $\sigma_0 = 5,25,70,100,200,500$ km/s     &                                                                                                                                                                                                     \\ \hline
\multirow{3}{*}{Morphology test (Sec. \ref{sec:morph-test})} & $n = 0.5, 1, 2, 4, 6, 8$              & \multirow{3}{*}{\begin{tabular}[c]{@{}c@{}}$\rm S/N = 10$, $i = 60^{\circ}$, $\rm PA = 45^{\circ}$,\\ $V_a = 200$ km/s, $\sigma_0 = 100$ km/s\end{tabular}}                                             \\ \cline{2-2}
                                 & $r_t = 0.25, 0.5, 1, 1.25, 1.5$       &                                                                                                                                                                                                     \\ \cline{2-2}
                                 & $\reff = 4.19 r_t$                      &                                                                                                                                                                                                     \\ \hline
Prior test - $n$ (Sec. \ref{sec:morph-test})      & $\Delta n_{\rm prior} = -0.5, 1,2,3$  & \begin{tabular}[c]{@{}c@{}}$\rm S/N = 10$, $i = 60^{\circ}$, $\rm PA = 45^{\circ}$,\\ $V_a = 200$ km/s, $\sigma_0 = 100$ km/s,\\ $r_{\rm e} = 4.19$, $n =1$, $\Delta r_{\rm e, prior} =  0$\end{tabular} \\ \hline
Prior test - $\reff$ (Sec. \ref{sec:morph-test})     & $\Delta r_{\rm e, prior} = -2,-1,1,2$ & \begin{tabular}[c]{@{}c@{}}$\rm S/N = 10$, $i = 60^{\circ}$, $\rm PA = 45^{\circ}$,\\ $V_a = 200$ km/s, $\sigma_0 = 100$ km/s,\\ $r_{\rm e} = 4.19$, $n =1$, $\Delta n_{\rm prior} = 0$\end{tabular} \\ \hline
\multirow{2}{*}{S/N test (Sec. \ref{sec:sn-test})}        & $\rm PA = 0,45,75^{\circ}$            & \multirow{2}{*}{\begin{tabular}[c]{@{}c@{}}$i = 60^{\circ}$, $V_a = 200$ km/s, $\sigma_0 = 100$ km/s,\\ $r_t = 1$, $r_{\rm e} = 4.19$, $n=1$\end{tabular}}                                          \\ \cline{2-2}
                                 & $\rm S/N = 3,5,10,20,50,100,200$      &                                                                                                                                                                                                     \\ 
\end{tabular}
\caption{Summary of the tests done on realistic mock data and presented in Sec. \ref{sec:real-tests}.}
\label{tab:tests}
\end{table*}




\bsp	
\label{lastpage}
\end{document}